\documentclass[submission,copyright,creativecommons]{eptcs}
\providecommand{\event}{AFL 2026} 

\usepackage{iftex}

\ifpdf
  \usepackage{underscore}         
  \usepackage[T1]{fontenc}        
\else
  \usepackage{breakurl}           
\fi

\usepackage{amssymb}
\usepackage{amsthm}
\usepackage{latexsym}
\usepackage[all]{xy}
\usepackage[american]{babel}
\usepackage{amsmath}
\usepackage{textcomp}

\newcommand{\TE}{\mathit{TR}}

\newcommand{\TC}{\mathit{TC}}
\newcommand{\TR}{\mathit{TR}}
\newcommand{\RConf}{\mathit{RConf}}

\usepackage{tikz}
\usepackage{marvosym}
\usetikzlibrary{shapes}
\usetikzlibrary{arrows,arrows.meta}
\usetikzlibrary{calc,through,backgrounds}
\usetikzlibrary{shapes.geometric,fit}

        \newtheorem{theorem}{Theorem}
        \newtheorem{proposition}{Proposition}
        \newtheorem{lemma}{Lemma}

         \newtheorem{definition}{Definition}

\newcounter{example}
\newenvironment{example}[1][]{\refstepcounter{example}\par\medskip
   \noindent \textit{Example~\theexample. #1} \rmfamily}{\medskip}

\title{Transition Systems from\\ Causal Reversible Bundle Event Structures}
\author{Nataliya Gribovskaya \qquad Irina Virbitskaite
\institute{A.P. Ershov Institute of Informatics Systems\\
the Siberian Branch of the Russian Academy of Sciences\\
6, Acad. Lavrentiev avenue,  630090, Novosibirsk, Russia}
\email{\{natamosk,virbitskaite\}@gmail.com}
}
\def\titlerunning{Transition Systems from Causal RBES}
\def\authorrunning{N. Gribovskaya, I. Virbitskaite}
\begin{document}
\maketitle

\begin{abstract}
Reversible computing is a novel paradigm that has recently emerged and extends traditional forwards-only computation with the capability to execute in the reverse direction,
making it possible for computation to run backwards as well as forwards.
Event structures are a foundational model in concurrency theory, providing a way to understand computational processes by describing the events that occur and the relationships between them.
In the literature, two structurally different approaches to associating transition system semantics with event structure models have been distinguished.
One approach is based on configurations, which are sets of already executed events.
The other approach is based on model residuals, which are not yet executed fragments of the model.
Configuration-based transition systems appear to be primarily used for semantic representations.
Residual-based transition systems are actively applied to demonstrate the consistency between operational and denotational semantics of concurrent process calculi,
as well as to visualize the dynamics of models.
The present paper focuses on bundle event structures with causal reversibility, which are used in the study of reversible extensions of CCS- and $\pi$-like systems.
Mappings from the reversible event structure model to the two transition system semantics are developed, which made it possible to prove the isomorphism of the semantics.
A category-theoretic characterization of the mappings is provided.

\end{abstract}

\section{Introduction}
The reversibility paradigm allows one to study computations that can proceed both in the standard, forward direction, and backward, returning to past states.
The notion of reversibility is natural in reliable systems since when an error occurs the system tries to go back to a previous consistent state.
Reversible computations in concurrent/distributed systems have many promising applications in software, hardware, bio-systems, and quantum computing.

Event structure models have become
crucial in the exploration of fundamental aspects of concurrent and nondeterministic computations,
including causality, conflict, and independence, and have found applications across a broad range of languages and models.
The connection between event structures and transition systems has been found to be beneficial for investigating and solving diverse problems in the analysis and verification of concurrent systems.
There are two major approaches to providing transition system semantics for event structures: configuration-based and residual-based.
In the first method (see \cite{K96,GP09,W89} among others),
states are understood as sets of events, called configurations,
and state transitions are built by starting with the empty configuration and enlarging configurations by already executed events.
In second more `structural' method (see \cite{BM94,BC89,CVY12,K96,L93} among others),
states are understood as event structures,
and transitions are created by starting with a given event structure as the initial state and then removing already executed and conflicting parts in the course of an execution.
In \cite{MR98}, the authors established that there is a close relationship in terms of bisimulation between these types of transition systems from prime event structures
within the interleaving/partial order spectrum.
The result of \cite{MR98} has been extended in \cite{BGV17} to more complex event structure models with asymmetric conflict.
The paper \cite{BGV18} demonstrated that isomorphisms, rather than bisimulations, between the two types of transition systems belonging to a single event structure can be obtained
for various event-based models and for a full range of semantics (interleaving, step, pomset, multiset).

Event structures have been expanded to represent reversible computational processes capable of undoing executed actions by allowing configurations to evolve by adding/removing events.
Reversible bundle event structures (RBESs) \cite{GPY21} are a concurrent and non-deterministic model in which some events are classified as reversible,
causal dependencies (`bundles' of events causing another events) and conflicts between events are explicitly presented, providing a non-interleaving semantics for reversible concurrent computations.
These models are more expressive than reversible prime event structures \cite{PU15,UPY18}, since they allow an event to have multiple possible conflicting causes,
thereby making it possible to model parallel composition without creating multiple copies of events.
RBESs are able to model a feature of reversible computing known as causal-consistent reversibility which relates reversibility to causality:
an event can be undone provided that all of its effects have been undone.
Causal RBESs are utilized in the study of reversible computations and controlled reversibility in CCS-like \cite{GPY21} and $\pi$-like systems \cite{GPY22},
and lay the foundation for adding rollback to the reversible process calculi, which cannot be done using reversible prime event structures.

The aim of this paper is to identify configuration- and residual-based types of transition system semantics for RBESs
and to understand how these semantics relate to each other, which can assist in the construction of algebraic calculi to describe reversible concurrent processes.
In \cite{GV23,GV25}, pairwise bisimilarity between the transition system semantics for cause-respecting reversible prime event structures with symmetric/asymmetric conflict was proved.
The present paper demonstrates that an isomorphism, rather than just a bisimulation, between the two types of transition systems from a causal RBES can be obtained.
A key idea is to employ non-executable (impossible) events \footnote{In an event structure, an event is called non-executable or impossible if it does not
occur in any configuration of the structure, i.e. the event is never executed.} that are allowed by the model in question, as opposed to (reversible) prime event structures.

This paper is structured as follows.
In Section 2, we start with recalling the syntax and semantics of bundle event structures (BESs),
continue with defining reversible bundle event structures (RBESs) and their (step) semantics in terms of (reachable) configurations,
and finish with considering reversing disciplines for the model under consideration.
In Section 3, we define a removal operator, which is useful for constructing RBES residuals, and demonstrate the correctness of the operator.
In Section 4, we develop the transition system semantics based on configurations and residuals of RBESs, and establish isomorphism results between the semantics in the context of causal RBESs.
In Section 5, we formulate a category-theoretic characterization of the mappings associating the transition systems with causal RBESs.
In Section 6, we provide some concluding remarks.
The proofs of the results presented here can be found at https://www.iis.nsk.su/AFL-2026.pdf.

\section{Reversing in Bundle Event Structures}
\subsection{Bundle Event Structures (BESs)}

Bundle event structures were introduced in \cite{L93} for the description of formal semantics of the specification language LOTOS for parallel systems
and the corresponding algebra of processes PA \cite{K96}.
In addition, the models have been used to provide semantics for other process algebras, including CSP, ACP, and CCS.
Unlike events in prime structures, those in bundle structures can be initiated by different sets of events.
Causality is not a binary relation anymore; instead, it is represented by the bundle relation $\mapsto$ between a finite set $W$ of events and event $e$.
A pair $(W, e)$ such that $W\mapsto e$ is called a bundle, and $W$ is called a bundle set.
The bundle set contains only pairwise conflicting events (which is known as the stability principle), ensuring unique enabling within an execution.
This causality relation can be interpreted as follows:
in the system's functioning, an event $e$ can occur only if one of the events from the set $W$ has already occurred.

\begin{definition}\label{def_BES}
A (labeled) {\em bundle event structure} (over a set $L$ of actions) (BES) is a 4-tuple
$\mathsf{E}=(E, \sharp, {\mapsto,} l)$,
where
$E$ is a set of events;
$\sharp \subseteq E \times E$ is an irreflexive and symmetric relation (the {\em conflict relation});
$ \mapsto \ \subseteq 2^E \times E $ is the {\em bundle relation},
satisfying
$W \mapsto e$ $\Rightarrow$ $\forall e_1, e_2 \in W \colon$ if $e_1\neq e_2$ then $e_1\ \sharp\ e_2$;
$l:E \to L$  is a labeling function.
\end{definition}

The behavior of the BES is described by explaining which subsets of events constitute possible (partial) runs of the represented system
(thus for\-mal\-ising the interpretation of the bundle sets and the conflict relation).
These subsets are called {\em configurations}.
In other words, a set $C\subseteq E$ is a {\em configuration} of the BES $\mathsf{E}$ iff
$C$ is {\em finite}; {\em conflict-free}, i.e., $\neg (e\ \sharp\ e')$, for all $e,e'\in C$; and {\em secured w.r.t. $\mapsto$},
i.e., there exist events $e_1, \ldots, e_n$ $(n\geq 0)$ such that $C=\{e_1, \ldots, e_n\}$, and
for all $0\leq i<n$, if $W \mapsto e_{i+1}$, then $\{e_1, \ldots, e_i\} \cap W \neq \emptyset$.
The causal relation between events within a configuration $C$ of the BES can be represented by the partial order
$\leq_C=\{(d,e)\in C\times C \mid \exists W \colon d\in W \mapsto e\}^\star$ ($^\star$ is the Kleene star).
The set of configurations of $\mathsf{E}$ is denoted by $Conf(\mathsf{E})$.

The definition of the syntax of the BES allows an empty bundle, $\emptyset\mapsto e$, to be defined.
The behavioral interpretation of such a bundle is that $e$ cannot occur in any configuration, i.e. $e$ is an impossible event.
Notice that there are alternative ways to specify impossible events, for instance $\{e\}\mapsto e$ or $\{e'\}\mapsto e\ \sharp\ e'$.
It is known from \cite{L93,K96} that all the bundles with impossible events can always be eliminated while preserving the behavior (in terms of configurations).

\subsection{Reversible Bundle Event Structures (RBESs)}

In this subsection, we recall the definition of reversible bundle event structures (RBESs) from \cite{GPY21,GPY22}.
In the reversible version of BESs,
some events are categorized as reversible each of whose are represented by $\underline{e}$,
the bundle relation extends to encompass reverse events,
a prevention relation is added,
such that if $e\rhd\underline{e}$ then $\underline{e}'$ cannot be reversed from configurations containing $e$.
We use $e^\ast$ to denote either $e$ or $\underline{e}$.
\begin{definition}\label{def_RBES}
A (labeled) {\em reversible bundle event structure (over a set $L$ of actions) (RBES)} is a 7-tuple
$\mathcal{E}=(E, F, \sharp, \mbox{$\mapsto$,} \rhd, l,C_0)$,
where
$E$ is the set of {\em events};
$F\subseteq E$ are those events of $E$ which are {\em reversible}, with reverse events being denoted by
$\underline{F}=\{\underline{e}\mid e\in F\}$;
$\sharp \subseteq E \times E$ is an irreflexive and symmetric relation (the {\em conflict relation});
$ \mapsto \ \subseteq 2^E \times (E\cup\underline{F}) $ is the {\em bundle relation},
satisfying
$W \mapsto e^\ast$ $\Rightarrow$ $\forall e_1, e_2 \in W \colon$
if $e_1\neq e_2$ then $e_1\ \sharp\ e_2$, and
$\forall e\in F\colon\{e\} \mapsto \underline{e}$;
$\rhd \subseteq E \times \underline{F}$ is the {\em prevention relation};
$l:E \to L$  is a labeling function;
$C_0\subseteq F$ is the initial configuration, which is finite, conflict-free, and secured w.r.t. $\mapsto\upharpoonright_{2^E \times E}$.
\end{definition}

The above definition heavily relies on Definition 3.15 from \cite{GPY21}.
Unlike Definition 3.15,
we enhance the capabilities of the RBES by starting its function with an initial configuration that is not necessarily an empty set.
In addition, we consider a labeled version of the RBES, which allows us to work with the models
that take into account auto-concurrency, auto-conflict, and auto-causality,
and demonstrate that the results obtained below are also apply to these models.

Notice that despite the fact that RBESs that include reversible prime event structures and are part of reversible stable event structures,
correspond to the same parameters in the taxonomy proposed in \cite{GLMMPUV23}.

For the graphical representation of RBESs, we use the following conventions.
We draw irreversible events in squares and reversible events in circles.
The bundles $(W,e)$ are indicated by drawing an arrow from each element of $W$ to $e$ and connecting all the arrows by lines;
the pairs of the events included in the conflict relation are marked by the symbol $\sharp$.
The reversible events belonging to the initial configuration are colored dark grey.
Clearly, if the initial configuration is an empty set then no events are colored dark grey.
For simplicity, in the following figures,
we will not depict the bundles of the form $\{e\}\mapsto\underline{e}$ since they exist for any $e\in F$.

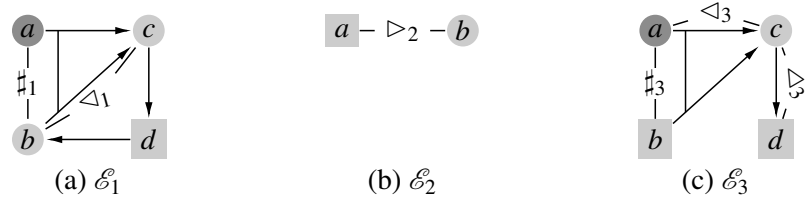
\begin{figure}[htbp]
\begin{center}
\begin{tikzpicture}[line width=0.02cm,>={Latex[length=0.22cm,width=0.1cm]},scale=0.8]
\filldraw[fill=gray!90,color=gray!90](0,2.5) circle (0.25);
\filldraw[fill=gray!40, color=gray!40](0,0.7) circle (0.25);
\filldraw[fill=gray!40, color=gray!40](2,2.5) circle (0.25);
\node[at={(0,2.5)}](a){$a$};
\node[at={(0,0.7)}](b){$b$};
\node[at={(2,2.5)}](c){$c$};
\node[fill=gray!40,at={(2,0.7)}](d){$d$};
\node[at={(1.,0)}](){(a) $\mathcal{E}_1$};

\node[at={(1.1,1.35)},rotate = 15](rhd){$\lhd_{1}$};
\draw[](c)edge(rhd);
\draw[](rhd)edge(b);

\path[->](b)edge[]node[]{}(c);
\path[->](a)edge[]node[]{}(c);
\path[->](c)edge[]node[]{}(d);
\path[->](d)edge[draw]node[]{}(b);
\draw[](b)edge[draw]node[fill=white,inner sep=1pt]{${\sharp_1}$}(a);
\draw (0.5,1.15) -- (0.5,2.5);
\end{tikzpicture}
\hspace*{2cm}\begin{tikzpicture}[line width=0.02cm,>={Latex[length=0.22cm,width=0.1cm]},scale=0.8]
\filldraw[fill=gray!40, color=gray!40](2,2.5) circle (0.25);
\node[fill=gray!40,at={(0,2.5)}](a){$a$};
\node[at={(2,2.5)}](b){$b$};

\node[at={(1.,2.5)},rotate = 0](rhd){$\rhd_{2}$};
\draw[](a)edge(rhd);
\draw[](rhd)edge(b);

\node[at={(1.,0)}](){(b) $\mathcal{E}_2$};

\end{tikzpicture}
\hspace*{2cm}\begin{tikzpicture}[line width=0.02cm,>={Latex[length=0.22cm,width=0.1cm]},scale=0.8]
\filldraw[fill=gray!90,color=gray!90](0,2.5) circle (0.25);
\filldraw[fill=gray!40, color=gray!40](2,2.5) circle (0.25);
\node[at={(0,2.5)}](a){$a$};
\node[fill=gray!40,at={(0,0.7)}](b){$b$};
\node[at={(2,2.5)}](c){$c$};
\node[fill=gray!40,at={(2,0.7)}](d){$d$};

\node[at={(1,2.8)},rotate = 0](rhd1){$\lhd_{3}$};
\draw[](c)edge(rhd1);
\draw[](rhd1)edge(a);

\node[at={(2.3,1.65)},rotate = -60](rhd2){$\lhd_{3}$};
\draw[](d)edge(rhd2);
\draw[](rhd2)edge(c);

\path[->](b)edge[]node[]{}(c);
\path[->](a)edge[]node[]{}(c);
\path[->](c)edge[]node[]{}(d);
\draw[](b)edge[draw]node[fill=white,inner sep=1pt]{${\sharp_3}$}(a);
\draw (0.5,1.15) -- (0.5,2.5);
\node[at={(1.,0)}](){(c) $\mathcal{E}_3$};
\end{tikzpicture}
\end{center}
\caption{Reversible bundle event structures}\label{R_Bund}
\end{figure}

\begin{example}\label{1.examp}\footnote{For simplicity, in this and the following examples,
we assume that the set of events is equal to the set of actions in each RBES, and the labeling function is identity, so its specification is omitted.}
Consider the structure $\mathcal{E}_1=(E_1$, $F_1$, $\sharp_1$, \mbox{$\mapsto_1$,} $\rhd_1$, $l_1$, $C^1_0)$ depicted in Fig.~\ref{R_Bund}(a).
Here,
$E_1=\{a,b,c,d\}$; $F_1=\{a,b,c\}$; $\sharp_1 = \{(a, b), (b,a)\}$;
$\mapsto_1 = \{(\{a\}, \underline{a})$, $(\{b\}, \underline{b})$, $(\{c\}, \underline{c})$, $(\{a,b\},c)$, $(\{c\},d)$, $(\{d\},b)\}$;
$\rhd_1 = \{(c,\underline{b})\}$;
$C^1_0 = \{a\}$.
It is easy to make sure that the components of the structure $\mathcal{E}_1$ meet the requirements of the corresponding items in Definition~\ref{def_RBES}.
We emphasize that the initial configuration $C^1_0$ contains the reversible event $a$.
So, the structure $\mathcal{E}_1$ is indeed an RBES.
\hfill$\Diamond$
\end{example}

We want to model the behavior of RBESs as configuration systems (CSs) \cite{PU15}.
A CS is a set of configurations, which are sets of events, some of these events are reversible, and there are transitions between configurations.
These configurations and transitions are described in the definition below.
\begin{definition}\label{def_CS}
A {\em configuration system} (CS) is a 4-tuple $CS=(E,F,\mathbb{C},\rightarrow)$,
where $E$ is a set of events;
$F\subseteq E$ is the set of {\em reversible events};
$\mathbb{C}\subseteq 2^E$ is a set of configurations; and
$\rightarrow\subseteq \mathbb{C}\times 2^{(E\cup\underline{F})}\times\mathbb{C}$ is a transition relation such that
if $C \xrightarrow{A\cup\underline{B}} C'$ then:
\begin{itemize}
\item $C,C'\in\mathbb{C}$, $C\cap A=\emptyset$, $B\subseteq C\cap F$, $C'=(C\setminus B)\cup A$;
\item for all $A'\subseteq A$ and $B'\subseteq B$,
we have $C \xrightarrow{A'\cup\underline{B'}} C'' \xrightarrow{(A\setminus A')\cup\underline{(B\setminus B')}} C'$,
meaning $C''=((C\setminus B')\cup A')\in\mathbb{C}$.
\end{itemize}
\end {definition}

Construct a CS as a representation of the behavior of the RBES as follows.
\begin{definition}\label{def_CS_of_RBES}
Define a mapping $C_{br}$ from RBESs to CSs as follows:\\
$C_{br}(\mathcal{E}=(E, F, \sharp, \mbox{$\mapsto$,} \rhd, l,C_0))=(E,F,\mathbb{C},\rightarrow)$, where:
\begin{itemize}
\item
$C\in\mathbb{C}$ if $C$ is finite and conflict-free;
\item
for $C\in\mathbb{C}$, $A\subseteq E$, and $B\subseteq F$, there is a transition
$C\stackrel{A \cup \underline{B}}{\longrightarrow}C'$ if $C'=((C\setminus B)\cup A)\in\mathbb{C}$, and
$A \cup \underline{B}$ is a {\em computation step enabled} at $C$, i.e. the following holds:
\begin{itemize}
\item[(a)] $A\cap C=\emptyset$, $B \subseteq C$, and $(C \cup A)$ is conflict-free;
\item[(b)] for all $e \in A$ and $W\subseteq E$, if $W\mapsto e$ then $W\cap (C \setminus B)\neq\emptyset$;
\item[(c)] for all $e \in B$ and $W\subseteq E$, if $W\mapsto\underline{e}$ then $W\cap (C \setminus (B\setminus\{e\}))\neq\emptyset$;
\item[(d)] for all $e \in B$, if $e' \rhd \underline{e}$ then $e' \not\in (C \cup A)$.
\end{itemize}
\end{itemize}
\end{definition}

In this way, RBESs progress by executing computation steps, i.e. executing events and/or undoing previously executed events, thus moving from one configuration to another.

Define the notion of reachable configurations for RBESs, which are finite subsets of non-conflicting events
that can be reached from the initial configuration by executing computation steps.

\begin{definition}\label{def_conf}
Given an RBES $\mathcal{E}=(E, F, \sharp, \mbox{$\mapsto$,} \rhd, l,C_0)$ and
its CS $C_{br}(\mathcal{E}=(E,F,\mathbb{C},\rightarrow))$,
$C\in\mathbb{C}$ is a {\em reachable (from $C_0$) configuration of $\mathcal{E}$} iff
for any $i = 1, \ldots, n$ $(n\geq0)$,
$C_{i-1} \stackrel{A_i\cup \underline{B}_i} \longrightarrow  C_{i}$ in $C_{br}(\mathcal{E})$, and $C_n=C$.
The set of reachable configurations of $\mathcal{E}$ is denoted by $RConf(\mathcal{E})$.
\end{definition}

\begin{figure}[htbp]
\begin{center}
\begin{tikzpicture}[line width=0.02cm,>={Latex[length=0.22cm,width=0.1cm]},scale=0.85]
\node[fill=blue!20,at={(1,4)}](E){$\emptyset$};
\node[fill=blue!20,at={(1,6.3)}](E-a){$\{a\}$};
\node[fill=blue!20,at={(4,4)}](E-c){$\{c\}$};
\node[fill=blue!20,at={(4,6.3)}](E-ac){$\{a,c\}$};
\node[fill=blue!20,at={(7,4)}](E-cd){$\{c,d\}$};
\node[fill=blue!20,at={(10,6.3)}](E-ad){$\{a,d\}$};
\node[fill=blue!20,at={(7,6.3)}](E-acd){$\{a,c,d\}$};
\node[fill=blue!20,at={(10,4)}](E-d){$\{d\}$};
\node[fill=blue!20,at={(13,4)}](E-bd){$\{b,d\}$};
\node[fill=blue!20,at={(10,1.7)}](E-bcd){$\{b,c,d\}$};

\path[-latex]([xshift=-0.13 cm]E.north)edge[]node[auto,sloped,above,inner sep=0.03cm,pos=0.5,rotate=0]{\small{$(\{a\}\cup\emptyset)$}}([xshift=-0.13 cm]E-a.south);
\path[-latex]([xshift=0.13 cm]E-a.south)edge[]node[auto,sloped,below,inner sep=0.03cm,pos=0.5,rotate=180]{\small{$(\emptyset \cup\{\underline{a}\})$}}([xshift=0.13 cm]E.north);

\path[-latex]([yshift=0.13 cm]E-a.east)edge[]node[auto,sloped,above,inner sep=0.03cm,pos=0.5,rotate=0]{\small{$(\{c\}\cup\emptyset)$}}([yshift=0.13cm]E-ac.west);
\path[-latex]([yshift=-0.13 cm]E-ac.west)edge[]node[auto,sloped,below,inner sep=0.03cm,pos=0.5,rotate=0]{\small{$(\emptyset\cup\{\underline{c}\})$}}([yshift=-0.13cm]E-a.east);

\path[-latex]([yshift=-0. cm]E-c.east)edge[]node[auto,sloped,below,inner sep=0.03cm,pos=0.5,rotate=0]{\small{$(\{d\}\cup\emptyset)$}}([yshift=-0. cm]E-cd.west);
\path[-latex]([yshift=-0.1 cm]E-ac.east)edge[]node[auto,sloped,above,inner sep=0.03cm,pos=0.5,rotate=0]{\small{$(\{d\}\cup\emptyset)$}}([yshift=-0.1 cm]E-acd.west);

\path[-latex]([xshift=-0.13 cm]E-c.north)edge[]node[auto,sloped,above,inner sep=0.03cm,pos=0.5,rotate=0]{\small{$(\{a\}\cup\emptyset)$}}([xshift=-0.13cm]E-ac.south);
\path[-latex]([xshift=0.13 cm]E-ac.south)edge[]node[auto,sloped,above,inner sep=0.03cm,pos=0.5,rotate=0]{\small{$(\emptyset\cup\{\underline{a}\})$}}([xshift=0.13cm]E-c.north);

\path[-latex]([xshift=-0.13 cm]E-cd.north)edge[]node[auto,sloped,above,inner sep=0.03cm,pos=0.5,rotate=0]{\small{$(\{a\}\cup\emptyset)$}}([xshift=-0.13cm]E-acd.south);
\path[-latex]([xshift=0.13 cm]E-acd.south)edge[]node[auto,sloped,above,inner sep=0.03cm,pos=0.5,rotate=0]{\small{$(\emptyset\cup\{\underline{a}\})$}}([xshift=0.13cm]E-cd.north);

\path[-latex]([yshift=-0.13 cm]E-ad.west)edge[]node[auto,sloped,below,inner sep=0.03cm,pos=0.5,rotate=0]{\small{$(\{c\}\cup\emptyset)$}}([yshift=-0.13cm]E-acd.east);
\path[-latex]([yshift=0.13 cm]E-acd.east)edge[]node[auto,sloped,above,inner sep=0.03cm,pos=0.5,rotate=0]{\small{$(\emptyset\cup\{\underline{c}\})$}}([yshift=0.13cm]E-ad.west);

\path[-latex]([xshift=-0.13 cm]E-d.north)edge[]node[auto,sloped,above,inner sep=0.03cm,pos=0.5,rotate=0]{\small{$(\{a\}\cup\emptyset)$}}([xshift=-0.13cm]E-ad.south);
\path[-latex]([xshift=0.13 cm]E-ad.south)edge[]node[auto,sloped,above,inner sep=0.03cm,pos=0.5,rotate=0]{\small{$(\emptyset\cup\{\underline{a}\})$}}([xshift=0.13cm]E-d.north);

\path[-latex]([yshift=-0.13 cm]E-bd.west)edge[]node[auto,sloped,below,inner sep=0.03cm,pos=0.5,rotate=0]{\small{$(\emptyset\cup\{\underline{b}\})$}}([yshift=-0.13cm]E-d.east);
\path[-latex]([yshift=0.13 cm]E-d.east)edge[]node[auto,sloped,above,inner sep=0.03cm,pos=0.5,rotate=0]{\small{$(\{b\}\cup\emptyset)$}}([yshift=0.13cm]E-bd.west);

\path[-latex]([yshift=-0. cm]E-cd.south)edge[]node[auto,sloped,above,inner sep=0.03cm,pos=0.5,rotate=0]{\small{$(\{b\}\cup\emptyset)$}}([xshift=-0.13cm]E-bcd.north);
\path[-latex]([xshift=-0.2 cm,yshift=-0. cm]E-bd.south)edge[]node[auto,sloped,above,inner sep=0.03cm,pos=0.5,rotate=0]{\small{$\!\!\!\!\!\!\!(\{c\}\cup\emptyset)$}}([xshift=-0.13cm]E-bcd.north);
\path[-latex]([xshift=0.2 cm,yshift=-0. cm]E-bcd.north)edge[]node[auto,sloped,below,inner sep=0.03cm,pos=0.5,rotate=0]{\small{$(\emptyset\cup\{\underline{c}\})$}}([xshift=0.13cm]E-bd.south);
\end{tikzpicture}
\caption{The behavior of $\mathcal{E}_1$ in the form of $C_{br}(\mathcal{E}_1)$}\label{Conf_E1}
\end{center}
\end{figure}
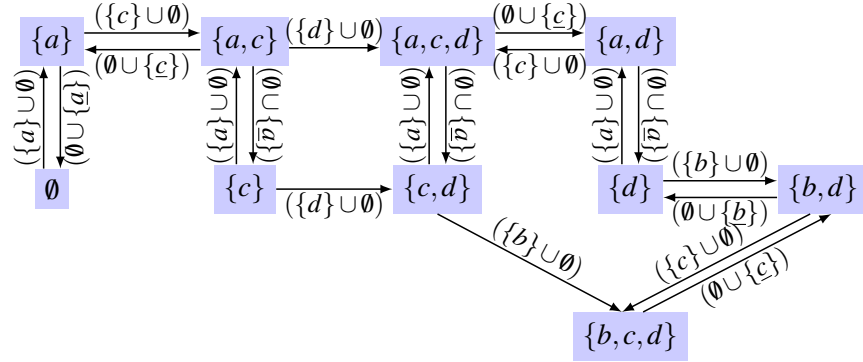

The example below illustrates the above definitions.
\begin{example}\label{2.examp}
Recall the RBES
$\mathcal{E}_1=(E_1$, $F_1$, $\sharp_1$, \mbox{$\mapsto_1$,} $\rhd_1$, $l_1$, $C^1_0)$ (see Example~\ref{1.examp} and Fig.~\ref{R_Bund}(a)). 
The behavior of $\mathcal{E}_1$ in the form of $C_{br}(\mathcal{E}_1)$ is shown in Fig.~\ref{Conf_E1}.
Using Definition~\ref{def_CS_of_RBES}, we observe the following.

According to item (a), in any computation step enabled at a configuration, the intersection of the $A$-part and the configuration is empty,
the $B$-part belongs to the configuration,
and the union of the $A$-part and the configuration does not contain conflicting events.
Notice that since $a\ \sharp_1\ b$, the events $a$ and $b$ cannot occur together in any configuration.

In accordance with item (b), the following applies to the $A$-parts in the computation steps.
The event $a$ can be executed at the configurations $\emptyset$, $\{c\}$, $\{d\}$, and $\{c,d\}$ because there is no bundle set $W\subseteq E_1$ such that $(W,a)\in\ \mapsto_1$.
Since the configurations $\{d\}$ and $\{c,d\}$ include $d$, the event $b$ can be executed at the configurations thanks to $(\{d\},b)\in\ \mapsto_1$.
Owing to $(\{a,b\},c)\in\ \mapsto_1$, the event $c$ can be executed at the configurations $\{a\}$, $\{a,d\}$, and $\{b,d\}$ because they all include either $a$ or $b$.
The event $d$ can be executed at the configurations $\{c\}$ and $\{a,c\}$, due to $(\{c\},d)\in\ \mapsto_1$.

Based on items (c) and (d), we obtain the following for the B-parts in the computation steps.
The event $a$ can be reversed at the configurations $\{a\}$, $\{a,c\}$, $\{a,d\}$, and $\{a,c,d\}$ because they all contain $a$ and there is no event in $\mathcal{E}_1$ which could prevent the undoing of $a$. 
As $\{b,d\}$ is the single configuration that contains $b$ and does not contain the event $c$ which prevents the reversal of $b$, the event $b$ can be undone only at $\{b,d\}$.
The event $c$ can be reversed at the configurations $\{a,c\}$, $\{a,c,d\}$, and $\{b,c,d\}$ because they all have $c$ 
and there is no event in $\mathcal{E}_1$ which could prevent the undoing of $c$.

Using Definition~\ref{def_conf} and Fig.~\ref{Conf_E1}, it is easy to see that in the RBES $\mathcal{E}_1$, its configurations coincide with its reachable configurations.
Notice that we can reach the configurations $\emptyset$, $\{c\}$, $\{d\}$, and $\{c,d\}$ from the initial configuration $\{a\}$ with a combination of forward and reverse steps
but we cannot reach them by doing only forward steps.
\hfill$\Diamond$
\end{example}

\subsection{Reversing Disciplines}

RBESs are able to model a feature of reversible computing known as causal-consistent reversibility which relates reversibility to causality:
an event can be undone provided that all of its effects have been undone.
This notion of reversibility is natural for reliable concurrent systems since the system attempts to return to a past consistent state, when an error occurs.
We recall Definition 3.31 from \cite{GPY21}.
\begin{definition}\label{def_cRBES}
An RBES
$\mathcal{E}=(E, F, \sharp, \mbox{$\mapsto$,} \rhd, l,C_0)$ is called:
\begin{itemize}
\item
{\em cause-respecting} (crRBES) iff whenever $W\mapsto e$ and $e'\in W\cap F$, then $e\rhd \underline{e'}$;
\item
{\em causal} (cRBES) iff:
\begin{itemize}
\item[(a)] it is cause-respecting,
\item[(b)]
whenever $W\mapsto\underline{e}$ then $e\in W$,
\item[(c)]
whenever $e\rhd\underline{e'}$ then either $e\ \sharp\ e'$ or there is $W\subseteq E$ such that $e'\in W\mapsto e$.
\end{itemize}
\end{itemize}
\end{definition}

Informally, in the cause-respecting RBES, effects of any reversible cause prevent it from being undone.
Then, if any reversible cause is undone at the current configuration, then its effects are absent there.
In addition, in the causal RBES, any reversible event is the only mandatory cause for its reversal,
and all events that prevent the reversal of the reversible event either conflict with it or are effects of it.
So, any reversible event can be undone at the current configuration only if it is present and their effects are absent.

\begin{example}\label{3.examp}
First, recall the RBES $\mathcal{E}_1$ (see Fig.~\ref{R_Bund}(a)) with the components:
$E_1=\{a,b,c,d\}$; $F_1=\{a,b,c\}$;
$\sharp_1$ $=$ $\{(a,b), (b,a)\}$;
$\mapsto_1$ $=$ $\{(\{a\}, \underline{a})$,  $(\{b\}, \underline{b})$, $(\{c\}, \underline{c})$, $(\{a,b\}, c)$, $(\{c\},d)$, $(\{d\},b)\}$;
$\rhd_1$ $=$ $\{(c, \underline{b})\}$;
$C^1_0$ $=$ $\{a\}$.
We see that $\{a,b\} \mapsto_1 c$ and $a\in F_1$.
However, the pair $(c, \underline{a})$ does not belong to the relation $\rhd_1$.
This means that the RBES is neither cause-respecting nor causal.

Second, consider the RBES $\mathcal{E}_2$ (see Fig.~\ref{R_Bund}(b)) with the components:
$E_2=\{a,b\}$; $F_2=\{b\}$;
$\sharp_2 = \emptyset$; $\mapsto_2 =  \{(\{b\}, \underline{b})\}$;
$\rhd_2 = \{(a, \underline{b})\}$;
$C^2_0 = \emptyset$.
As the relation $\mapsto_2$ does not contain any pair of the form $W\mapsto_2 e$, where $W\subseteq E_2$ and $e\in E_2$,
the RBES is cause-respecting.
On the other hand, $\mathcal{E}_2$ is not causal, because $a \rhd_2 \underline{b}$,
but $a$ and $b$ are not in the conflict relation, and, besides, there is no bundle set $W\subseteq E_2$ such that $b\in W \mapsto_2 a$.

Finally, observe the RBES $\mathcal{E}_3$ (see Fig.~\ref{R_Bund}(c)) with the components:
$E_3=\{a,b,c,d\}$; $F_3=\{a,c\}$;
$\sharp_3 = \{(a,b)$, $(b,a)\}$;
$\mapsto_3 = \{(\{a,b\},c)$, $(\{c\},d), (\{a\}, \underline{a}), (\{c\}, \underline{c})\}$;
$\rhd_3 = \{(c, \underline{a}), (d, \underline{c})\}$;
$C^3_0 = \{a\}$.
We see that $(\{a,b\},c)$, $(\{c\},d)\in\mapsto_3$, and $F_3=\{a,c\}$.
Since $c \rhd_3 \underline{a}$ and $d \rhd_3 \underline{c}$, the RBES is cause-respecting.
Clearly, item (c) in Definition~\ref{def_cRBES} holds.
In addition, item (b) in Definition~\ref{def_cRBES} is true,
as we have $(\{a\},\underline{a})$, $(\{c\},\underline{c})\in\mapsto_3$ for any event in $F_3$.
Therefore, $\mathcal{E}_3$ is a causal RBES.
\hfill$\Diamond$
\end{example}

The following lemma outlines specific features of cause-respecting RBESs.
\begin{lemma}\label{lem_secured}
Given a cause-respecting RBES
$\mathcal{E}$ and $C\in RConf(\mathcal{E})$,
$C$ is finite, conflict-free, and secured w.r.t. $\mapsto\upharpoonright_{2^E \times E}$.
\end{lemma}

Notice that for the RBES with the non-empty initial configuration,
Proposition 3.32 from \cite{GPY21} (which establishes that any reachable configuration is forward reachable)
does not hold even in the context of causal RBESs, as demonstrated at the end of the following example.

\begin{example}\label{4.examp}
Recall the non-cause-respecting RBES $\mathcal{E}_1$ (see Examples~\ref{1.examp}--\ref{3.examp}). 
From Example~\ref{2.examp}, we know that
the reachable configurations of $\mathcal{E}_1$ are: $\{a\}$, $\emptyset$, $\{a,c\}$, $\{a,c,d\}$, $\{c\}$, $\{c,d\}$, $\{a,d\}$, $\{d\}$, $\{b,d\}$, $\{b,c,d\}$.
Clearly, these configurations are finite and conflict-free.
However, it is easy to check that the configurations $\{c\}$, $\{c,d\}$, $\{a,d\}$, $\{d\}$, $\{b,d\}$, $\{b,c,d\}$ are not secured w.r.t. $\mapsto_1\upharpoonright_{2^{E_1} \times E_1}$.

Consider the cause-respecting RBES $\mathcal{E}_2$ from Example~\ref{3.examp}. 
As the conflict relation $\sharp_2$ is empty,
we have the following conflict-free and obviously finite configurations: $\emptyset$, $\{a\}$, $\{b\}$, and $\{a,b\}$.
According to Definitions~\ref{def_CS_of_RBES} and \ref{def_conf}, we get the following.
The events $a$ and $b$ can be executed in any order or together at the initial configuration $\emptyset$.
So, the reachable configurations of $\mathcal{E}_2$ are: $\emptyset$, $\{a\}$, $\{b\}$, and $\{a,b\}$.
Notice that the event $b$ can be canceled at the configuration $\{b\}$ (moving to $\emptyset$), but not at the configuration $\{a,b\}$ because $\rhd_2 = \{(a, \underline{b})\}$.
Since this RBES does not contain any bundle of the form $W\mapsto_2 e$, where $W\subseteq E_2$ and $e\in E_2$,
all the above reachable configurations are secured w.r.t. $\mapsto_2\upharpoonright_{2^{E_2} \times E_2}$.

Finally, observe the causal RBES $\mathcal{E}_3$ from Example~\ref{3.examp}. 
Using Definitions~\ref{def_CS_of_RBES} and \ref{def_conf}, we get the following.
As the event $a$ has the only pair $(\{a\}, \underline{a})$ in $\mapsto_3$ and the only pair $(c, \underline{a})$ in $\rhd_3$,
it can be reversed at the initial configuration $\{a\}$. 
Hence, the empty set is a reachable configuration of $\mathcal{E}_3$.
Notice that the events $a$ and $b$ are in conflict, and, hence, they cannot occur together in any configuration.
Since the event $a$ and $b$ have no bundle sets, they can be executed at the empty configuration.
The event $c$ can only be executed at the configurations that contain either $a$ or $b$, because  $(\{a,b\}, c)\in \mapsto_3$.
As $(\{c\}, d)\in\mapsto_3$, the event $d$ can only be executed after the event $c$ has been executed.
Therefore, we may conclude that
the reachable configurations of $\mathcal{E}_3$ are: $\{a\}$, $\emptyset$, $\{b\}$, $\{a,c\}$, $\{b,c\}$, $\{a,c,d\}$, $\{b,c,d\}$.
Due to $\rhd_3 = \{(c, \underline{a}), (d, \underline{c})\}$,
the event $a$ cannot be reversed at any configurations except for at the configuration $\{a\}$,
and the event $c$ can be reversed at the configurations $\{a,c\}$ and $\{b,c\}$.
It is easy to see that all the reachable configurations of $\mathcal{E}_3$ are finite, conflict-free, and secured w.r.t. $\mapsto_3\upharpoonright_{2^{E_3} \times E_3}$.
However, the configurations $\emptyset$, $\{b\}$, $\{b,c\}$, and  $\{b, c, d\}$ are not forward-reachable
from the initial configuration $\{a\}$, because the reversal of the event $a$ is needed.
\hfill$\Diamond$
\end{example}

Now we move on to defining a mapping from BESs to RBESs and demonstrating the similarity in the behavior of the models.
The fact below directly follows from the definitions given prior to that.

\begin{lemma}\label{BES_to_RBPES}
For any BES $\mathsf{E}=(E, \sharp, \mapsto, l)$,
$\alpha(\mathsf{E})=(E, \emptyset, \sharp, \mbox{$\mapsto$,} \emptyset, l,\emptyset)$ is a causal RBES such that $Conf(\mathsf{E})=RConf(\alpha(\mathsf{E}))$.
\end{lemma}

\section{Residuals of RBESs}
The removal operator, which is based on the idea of eliminating events that have already been executed in the reachable configuration and cannot be reversed subsequently,
as well as making events that conflict with those non-executable, is essential for constructing residual transition system semantics for the RBES.

We introduce the definition of the removal operator for the RBES.
\begin{definition}\label{def_rem}
For an RBES
$\mathcal{E}=(E, F, \sharp,\mapsto, \rhd, l,C_0)$ and a reachable configuration $C\in RConf(\mathcal{E})$,
the {\em residual} $\mathcal{E}\setminus C$ of $\mathcal{E}$ after $C$ under the removal operator $\setminus$ is defined as follows:
$\mathcal{E}\setminus C= (E'$, $F'$, $\sharp'=\sharp\cap(E'\times E')$, $\mapsto'$, $\rhd'=\rhd\cap\ (E'\times \underline{F}')$, $l'=l\mid_{E'}$, $C'_0=C\cap F')$,
with:
\begin{itemize}
\item
$E'=E\setminus\widetilde{C}$;
\item
$F'=\big(F\cap E'\big)\setminus\sharp\widetilde{C}$;
\item
$\mapsto'=\big(\!\!\mapsto \cap (2^{E'}\times{E'\cup\underline{F}'})\big) \cup \{(\{e\},e)\mid e\in\sharp\widetilde{C}\}$.
\end{itemize}
Here, $\widetilde{C}=\{e'\in C\mid \exists e\in(C\setminus F) \ \colon\ e'\leq_{C} e\}$ and $\sharp\widetilde{C}=\{e'\in E\mid \exists e\in \widetilde{C}\colon e'\ \sharp\ e\}$.
\end{definition}

The intuitive interpretation of the above definition is as follows.

In the residual, the set $E'$ of events is the result of subtracting the set $\widetilde{C}$ of the irreversible events, along with their causes, from the original set $E$.
This is because the events in $\widetilde{C}$ have already been executed in the reachable configuration $C$.

In the residual, the set $F'$ of reversible events is formed from the set $F$ by leaving the reversible events that remain in $E'$ and deleting the events from $\sharp\widetilde{C}$.
The reversible events in the set $\sharp\widetilde{C}$ do not need to be reversed because they will never be executed due to conflicts with events in the set $\widetilde{C}$.

In the residual, the bundle relation $\mapsto'$ is reduced by taking into account the events in the sets $E'$ and $\underline{F'}$.
Additionally, the relation is extended by adding the bundle set $\{e\}$ for all events $e\in\sharp\widetilde{C}$
(i.e., these events become non-executable) because they can no longer be executed.

The other components of the residual are defined on the irreversible and reversible events that remain in the residual.
\medskip

Consider illustrations of how the RBES operator acts.
\begin{example}\label{5.examp}
Consider the RBES $\mathcal{E}_3$ (see Examples~\ref{3.examp}--\ref{4.examp}) with the components:
$E_3=\{a,b,c,d\}$; $F_3=\{a,c\}$;
$\sharp_3 = \{(a,b)$, $(b,a)\}$;
$\mapsto_3 = \{(\{a,b\},c)$, $(\{c\},d), (\{a\}, \underline{a}), (\{c\}, \underline{c})\}$;
$\rhd_3 = \{(c, \underline{a}), (d, \underline{c})\}$;
$C^3_0 = \{a\}$.
From Example~\ref{4.examp}, we know that the reachable configurations of $\mathcal{E}_3$ are: $\{a\}$, $\emptyset$, $\{b\}$, $\{a,c\}$, $\{b,c\}$, $\{a,c,d\}$, $\{b,c,d\}$.
\medskip

Applying the removal operator to the RBES $\mathcal{E}_3$ and its reachable configurations,
we obtain the following structures:
\begin{itemize}
\item[--] $\mathcal{E}_3=\mathcal{E}_3\setminus\{a\}$ due to $a\in F_3$ (see Fig.~\ref{rem_RBES3}(a)).
\item[--] $\mathcal{E}_3\setminus \emptyset =
(\tilde{E}=E_3$, $\tilde{F}=F_3$, $\tilde{\sharp}=\sharp_3$, $\tilde{\mapsto}=\ \mapsto_3$, $\tilde{l}=l_3$, $\tilde{\rhd}=\rhd_3$, $\tilde{C}_0=\emptyset)$
because $\tilde{C}_0=\emptyset \cap\tilde{F}$ (see Fig.~\ref{rem_RBES3}(b));
\item[--]
$\mathcal{E}_3\setminus \{a,c\} =
(\dot{E}=E_3$, $\dot{F}=F_3$, $\dot{\sharp}=\sharp_3$, $\dot{\mapsto}\ =\ \mapsto_3$, $\dot{l}=l_3$, $\dot{\rhd}=\rhd_3$, $\dot{C}_0=\{a,c\})$,
as $a, c\in F_3$ and $\dot{C}_0=\{a,c\}\cap(\dot{F}=\{a,c\})$ (see Fig.~\ref{rem_RBES3}(c));
\item[--]
$\mathcal{E}_3\setminus \{a,c,d\} =
({\ddot{E}}=\{b\}$, ${\ddot{F}}=\emptyset$, ${\ddot{\sharp}}=\emptyset$, ${\ddot{\mapsto}}=\{(\{b\}, b)\}$, ${\ddot{l}}=l_3\mid_{\ddot{E}}$, ${\ddot{\rhd}}=\emptyset$, ${\ddot{C}}_0=\emptyset$),
thanks to $d\in(\{a,c,d\}\setminus F_3=\{a,c\})$, $a\leq_{\{a,c,d\}}c\leq_{\{a,c,d\}}d$, i.e. $\widetilde{\{a,c,d\}}=\{a,c,d\}$, $b\in\sharp_3\widetilde{\{a,c,d\}}$,
and ${\ddot{C}}_0=\{a,c,d\}\cap\ddot{F}$ (see Fig.~\ref{rem_RBES3}(d));
\item[--]
$\mathcal{E}_3\setminus \{b\}=
({\hat{E}}=\{a,c,d\}$, ${\hat{F}}=\{c\}$, ${\hat{\sharp}}=\emptyset$, ${\hat{\mapsto}}=\{(\{a\}, a), (\{c\}, d), (\{c\}, \underline{c})\}$, ${\hat{l}}=l_3\mid_{\hat{E}}$,
${\hat{\rhd}}=\{(d, \underline{c})\}$, ${\hat{C}}_0=\emptyset$), since $b\not\in {F_3}$, i.e. $\widetilde{\{b\}}=\{b\}$, $a\in\sharp_3\widetilde{\{b\}}$, and $\hat{C}_0=\{b\} \cap\hat{F}=\emptyset$
(see Fig.~\ref{rem_RBES3}(e));
\item[--]
$\mathcal{E}_3\setminus \{b,c\} =
(\breve{E}=\{a,c,d\}$, ${\breve{F}}=\{c\}$, $\breve{\sharp}=\emptyset$, $\breve{\mapsto}=\{(\{a\}, a), (\{c\}, d), (\{c\}, \underline{c})\}$, $\breve{l}=l_3|_{\breve{E}}$;
 $\breve{\rhd} = \{(d,\underline{c})\}$, $\breve{C}_0 = \{c\})$,
because $b\not\in{F_3}$ and $c\in{F_3}$, i.e. $\widetilde{\{b,c\}}=\{b\}$, $a\in\sharp_3\widetilde{\{b,c\}}$, and $\breve{C}_0=\{b,c\}\cap\breve{F}$ (see Fig.~\ref{rem_RBES3}(f));
\item[--]
$\mathcal{E}_3\setminus \{b,c,d\} =
({\check{E}}=\{a\}$, ${\check{F}}=\emptyset$, ${\check{\sharp}}=\emptyset$, ${\check{\mapsto}}=\{(\{a\}, a)\}$, ${\check{l}}=l_3\mid_{\check{E}}$,
 ${\check{\rhd}}=\emptyset$, ${\check{C}}_0=\emptyset$), due to
$b,d\in(\{b,c,d\}\setminus F_3=\{a,c\})$, $b\leq_{\{b,c,d\}}c\leq_{\{b,c,d\}}d$, i.e. $\widetilde{\{b,c,d\}}=\{b,c,d\}$, $a\in\sharp_3\widetilde{\{b,c,d\}}$, and
${\check{C}}_0=\{b,c,d\} \cap \check{F}$ (see Fig.~\ref{rem_RBES3}(g)).
\hfill$\Diamond$
\end{itemize}
\end{example}

\begin{figure}[htbp]\label{rem_RBES3}
\begin{center}
\begin{tikzpicture}[line width=0.02cm,>={Latex[length=0.22cm,width=0.1cm]},scale=0.8]
\filldraw[fill=gray!90,color=gray!90](0,2.5) circle (0.25);
\filldraw[fill=gray!40, color=gray!40](2,2.5) circle (0.25);
\node[at={(0,2.5)}](a){$a$};
\node[fill=gray!40,at={(0,0.7)}](b){$b$};
\node[at={(2,2.5)}](c){$c$};
\node[fill=gray!40,at={(2,0.7)}](d){$d$};
\node[at={(1,2.8)},rotate = 0](rhd1){$\lhd_{3}$};
\draw[](c)edge(rhd1);
\draw[](rhd1)edge(a);
\node[at={(2.3,1.6)},rotate = -60](rhd2){$\lhd_{3}$};
\draw[](d)edge(rhd2);
\draw[](rhd2)edge(c);
\path[->](b)edge[]node[]{}(c);
\path[->](a)edge[]node[]{}(c);
\path[->](c)edge[]node[]{}(d);
\draw[](b)edge[draw]node[fill=white,inner sep=1pt]{${\sharp_3}$}(a);
\draw (0.5,1.15) -- (0.5,2.5);
\node[at={(1.,-0.3)}](){(a) $\mathcal{E}_3=\mathcal{E}_3\setminus\{a\}$};
\end{tikzpicture}
\hspace*{1.cm}
\begin{tikzpicture}[line width=0.02cm,>={Latex[length=0.22cm,width=0.1cm]},scale=0.8]
\filldraw[fill=gray!40,color=gray!40](0,2.5) circle (0.25);
\filldraw[fill=gray!40, color=gray!40](2,2.5) circle (0.25);
\node[at={(0,2.5)}](a){$a$};
\node[fill=gray!40,at={(0,0.7)}](b){$b$};
\node[at={(2,2.5)}](c){$c$};
\node[fill=gray!40,at={(2,0.7)}](d){$d$};
\node[at={(1,2.8)},rotate = 0](rhd1){$\tilde{\lhd}$};
\draw[](c)edge(rhd1);
\draw[](rhd1)edge(a);
\node[at={(2.3,1.6)},rotate = -60](rhd2){$\tilde{\lhd}$};
\draw[](d)edge(rhd2);
\draw[](rhd2)edge(c);
\path[->](b)edge[]node[]{}(c);
\path[->](a)edge[]node[]{}(c);
\path[->](c)edge[]node[]{}(d);
\draw[](b)edge[draw]node[fill=white,inner sep=1pt]{$\tilde{\sharp}$}(a);
\draw (0.5,1.15) -- (0.5,2.5);
\node[at={(1.,-0.3)}](){(b) $\mathcal{E}_3\setminus\emptyset:$};
\end{tikzpicture}
\hspace*{1.cm}
\begin{tikzpicture}[line width=0.02cm,>={Latex[length=0.22cm,width=0.1cm]},scale=0.8]
\filldraw[fill=gray!90,color=gray!90](0,2.5) circle (0.25);
\filldraw[fill=gray!90, color=gray!90](2,2.5) circle (0.25);
\node[at={(0,2.5)}](a){$a$};
\node[fill=gray!40,at={(0,0.7)}](b){$b$};
\node[at={(2,2.5)}](c){$c$};
\node[fill=gray!40,at={(2,0.7)}](d){$d$};
\node[at={(1,2.8)},rotate = 0](rhd1){$\dot{\lhd}$};
\draw[](c)edge(rhd1);
\draw[](rhd1)edge(a);
\node[at={(2.3,1.6)},rotate = -60](rhd2){$\dot{\lhd}$};
\draw[](d)edge(rhd2);
\draw[](rhd2)edge(c);
\path[->](b)edge[]node[]{}(c);
\path[->](a)edge[]node[]{}(c);
\path[->](c)edge[]node[]{}(d);
\draw[](b)edge[draw]node[fill=white,inner sep=1pt]{$\dot{\sharp}$}(a);
\draw (0.5,1.15) -- (0.5,2.5);
\node[at={(1.,-0.3)}](){(c) $\mathcal{E}_3\setminus\{a,c\}$};
\end{tikzpicture}
\hspace*{1.cm}
\begin{tikzpicture}[line width=0.02cm,>={Latex[length=0.22cm,width=0.1cm]},scale=0.8]
\node[fill=gray!40,at={(1,2.5)}](b){$b$};
\draw[-latex](b) .. controls (1,1) and (2,2) ..
node[ellipse,sloped,above,inner sep=0pt,pos=0.5]{}(b);
\node[at={(1.,-0.3)}](){(d) $\mathcal{E}_3\setminus\{a,c,d\}$};
\end{tikzpicture}
\end{center}
\medskip
\begin{center}
\begin{tikzpicture}[line width=0.02cm,>={Latex[length=0.22cm,width=0.1cm]},scale=0.8]
\filldraw[fill=gray!40, color=gray!40](2,2.5) circle (0.25);
\node[fill=gray!40,at={(1,2.5)}](a){$a$};
\node[at={(2,2.5)}](c){$c$};
\node[fill=gray!40,at={(2,0.7)}](d){$d$};
\node[at={(2.3,1.6)},rotate = -60](rhd2){$\hat{\lhd}$};
\draw[](d)edge(rhd2);
\draw[](rhd2)edge(c);
\path[->](c)edge[]node[]{}(d);
\draw[-latex](a) .. controls (1,1) and (2,2) ..
node[ellipse,sloped,above,inner sep=0pt,pos=0.5]{}(a);
\node[at={(1.5,-0.3)}](){(e) $\mathcal{E}_3\setminus\{b\}$};
\end{tikzpicture}
\hspace*{1cm}
\begin{tikzpicture}[line width=0.02cm,>={Latex[length=0.22cm,width=0.1cm]},scale=0.8]
\filldraw[fill=gray!90, color=gray!90](2,2.5) circle (0.25);
\node[fill=gray!40,at={(1,2.5)}](a){$a$};
\node[at={(2,2.5)}](c){$c$};
\node[fill=gray!40,at={(2,0.7)}](d){$d$};
\node[at={(2.3,1.6)},rotate = -60](rhd2){$\breve{\lhd}$};
\draw[](d)edge(rhd2);
\draw[](rhd2)edge(c);
\path[->](c)edge[]node[]{}(d);
\draw[-latex](a) .. controls (1,1) and (2,2) .. node[ellipse,sloped,above,inner sep=0pt,pos=0.5]{}(a);
\node[at={(1.5,-0.3)}](){(f) $\mathcal{E}_3\setminus\{b,c\}$};
\end{tikzpicture}
\hspace*{1cm}
\begin{tikzpicture}[line width=0.02cm,>={Latex[length=0.22cm,width=0.1cm]},scale=0.8]
\node[fill=gray!40,at={(1,2.5)}](a){$a$};
\draw[-latex](a) .. controls (1,1) and (2,2) ..
node[ellipse,sloped,above,inner sep=0pt,pos=0.5]{}(a);
\node[at={(1.,-0.3)}](){(g) $\mathcal{E}_3\setminus\{b,c,d\}$};
\end{tikzpicture}
\end{center}
\caption{}
\end{figure}

Below are some specific facts related to the RBES removal operator.

\begin{lemma}\label{lem_rem_0}
Given a causal RBES $\mathcal{E}=(E, F, \sharp, \mapsto, \rhd, l, C_0)$, its reachable configuration $C \in RConf(\mathcal{E})$, and
the residual $\mathcal{E}\setminus C=(E', F', \sharp', \mapsto', \rhd', l', C'_0)$, it holds:
\begin{itemize}
\item[(i)]
$E'\subseteq E$, $F'\subseteq F$, $\sharp'\subseteq\sharp$, $\mapsto''\ \subseteq\ \mapsto$, 
$\rhd'\subseteq\rhd$, $C'_0\subseteq C$,
where $\mapsto''\ =\big(\!\!\mapsto \cap (2^{E'}\times{E'\cup\underline{F}'})\big)$;
\item[(ii)] $\mathcal{E}\setminus C_0 = \mathcal{E}$;
\item[(iii)] $\widetilde{C} \subseteq C$;
\item[(iv)] $\sharp\widetilde{C}\subseteq E'$;
\item[(v)] $C'_0=C\cap E'$;
\item[(vi)] $\mathcal{E}\setminus C$ is a causal RBES.
\end{itemize}
\end{lemma}

The following statement demonstrates compositional properties of the residual operator for causal RBESs.

\begin{proposition}\label{prop_1}
Given a causal RBES $\mathcal{E}=(E, F, \sharp, {\mapsto,} \rhd, l, C_0)$, a reachable configuration $C\in RConf(\mathcal{E})$,
and the residual $\mathcal{E}' = \mathcal{E}\setminus C=(E', F', \sharp', {\mapsto',} \rhd', l', C'_0)$,
$C \stackrel{A\cup \underline{B}}{\longrightarrow} C'$ in $\mathcal{E}$ iff
$C'_0 \stackrel{A\cup \underline{B}}{\longrightarrow'} C''$ in $\mathcal{E}'$, and, moreover, $\mathcal{E}\setminus C' = \mathcal{E}'\setminus C''$.
\end{proposition}

The meaning of the above properties lies in the fact that
the obtained residuals of causal RBESs do not allow reachable configurations
that are unreachable in the original structures.

\begin{figure}[htbp]
\begin{center}
\begin{tikzpicture}[line width=0.02cm,>={Latex[length=0.22cm,width=0.1cm]},scale=0.8]
\filldraw[fill=gray!40, color=gray!40](2,2.5) circle (0.25);
\node[fill=gray!40,at={(0,2.5)}](a){$a$};
\node[at={(2,2.5)}](b){$b$};

\node[at={(1.,2.5)},rotate = 0](rhd){$\rhd_{2}$};
\draw[](a)edge(rhd);
\draw[](rhd)edge(b);

\node[at={(1.,1)}](){(a) $\mathcal{E}_2=\mathcal{E}_2\setminus\emptyset$};

\end{tikzpicture}
\hspace*{1cm}\begin{tikzpicture}[line width=0.02cm,>={Latex[length=0.22cm,width=0.1cm]},scale=0.8]
\filldraw[fill=gray!40, color=gray!40](1,2.5) circle (0.25);
\node[at={(1,2.5)}](b){$b$};
\node[at={(1.,1)}](){(b) $\mathcal{E}_2\setminus\{a\}$};
\end{tikzpicture}
\hspace*{1cm}\begin{tikzpicture}[line width=0.02cm,>={Latex[length=0.22cm,width=0.1cm]},scale=0.8]
\filldraw[fill=gray!90, color=gray!90](2,2.5) circle (0.25);
\node[fill=gray!40,at={(0,2.5)}](a){$a$};
\node[at={(2,2.5)}](b){$b$};
\node[at={(1.,2.5)},rotate = 0](rhd){$\rhd_{2}$};
\draw[](a)edge(rhd);
\draw[](rhd)edge(b);
\node[at={(1.,1)}](){(c) $\mathcal{E}_2\setminus\{b\}$};
\end{tikzpicture}
\hspace*{1cm}\begin{tikzpicture}[line width=0.02cm,>={Latex[length=0.22cm,width=0.1cm]},scale=0.8]
\filldraw[fill=gray!90, color=gray!90](1,2.5) circle (0.25);
\node[at={(1,2.5)}](b){$b$};
\node[at={(1.,1)}](){(d) $\mathcal{E}_2\setminus\{a,b\}$};
\end{tikzpicture}
\end{center}
\caption{The residuals of the RBES $\mathcal{E}_2$}\label{rem_RBES2}
\end{figure}
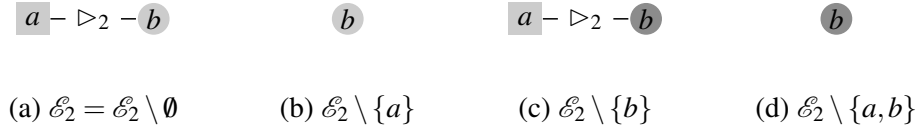

\begin{example}\label{7.examp}
First, examine the non-cause-respecting RBES $\mathcal{E}_1$ (see Examples~\ref{1.examp}--\ref{4.examp} and Fig.~\ref{R_Bund}(b)).
From Example~\ref{2.examp}, we know that $C =\{a,c,d\}\in RConf(\mathcal{E}_1)$ and
$\{a,c,d\}\stackrel{(\emptyset\cup\{\underline{a}\})}\rightarrow\{c,d\}\in RConf(\mathcal{E}_1)$.
Using Definition~\ref{def_rem}, we obtain the residual
$\mathcal{E}_1\setminus C = (E'_1=\{b\}$, $F'_1=\emptyset$, $\sharp'_1=\emptyset$, $\mapsto'_1=\{(\{b\},b)\}$, $l'_1 = l_1\mid_{E'_1}$, $\rhd'_1 = \emptyset$,
$C'^1_0 = \emptyset)$,
because $\widetilde{C} = C$ and $\sharp_1\widetilde{C} = \{b\}$.
According to Definition~\ref{def_CS_of_RBES}, the computation step $(\emptyset \cup \{\underline{a}\})$ is not enabled at $C'^1_0$.

Second, consider the cause-respecting and non-causal RBES $\mathcal{E}_2$ discussed in Examples~\ref{3.examp}--\ref{4.examp} and shown in Fig.~\ref{R_Bund}(b).
As demonstrated in Example~\ref{4.examp}, the reachable configurations of $\mathcal{E}_2$ are: $\emptyset$, $\{a\}$, $\{b\}$, $\{a,b\}$.
The residuals of $\mathcal{E}_2$ after the configurations are drawn in Fig.~\ref{rem_RBES2}.
Contemplate the residual $\mathcal{E}_2\setminus \{a,b\}$ depicted in Fig.~\ref{rem_RBES2}(d).
Using Definition~\ref{def_CS_of_RBES}, we get the following.
The computation step $(\emptyset \cup \{\underline{b}\})$ is enabled at the initial configuration $\{b\}$ of $\mathcal{E}_2\setminus \{a,b\}$,
whereas the computation step is not enabled at $\{a,b\}$ in $\mathcal{E}_2$, because $\{a,b\}$ contains the event $a$ preventing the reversal of $b$.

Using Examples~\ref{3.examp}--\ref{5.examp},
it is not difficult to check that Proposition~\ref{prop_1} holds for the causal RBES $\mathcal{E}_3$.
\hfill$\Diamond$
\end{example}

\section{Transition System Semantics for RBESs}\label{assoc.sct}

In this section,
we first give some basic definitions concerning labeled transition systems.
Then, we define the mappings $\TC(\mathcal{E})$ and $\TE(\mathcal{E})$,
which associate with the RBES $\mathcal{E}$ two distinct kinds of transition systems
-- one whose states are reachable configurations and one whose states are residuals of $\mathcal{E}$.

For the fixed set $L$ of actions in RBESs,
define the set $\mathbb{L}:= \mathbb{N}_0^{(L\cup \underline{L})}$ (the set of multisets over the set $(L\cup \underline{L})$). 
The set $\mathbb{L}$ will be used to represent a set of labels for transition systems.

A (labeled) transition system $T = (S,\rightarrow,i)$ (over a set $\mathbb{L}$ of labels) (LTS) consists of a set of states $S$, a transition relation $\rightarrow\subseteq S\times\mathbb{L}\times S$,
and an initial state $i\in S$.
Two labeled transition systems over $\mathbb{L}$ are {\em isomorphic} if their states can be mapped one-to-one to each other, preserving transitions and initial states.

We are ready to define labeled transition systems with reachable configurations as states.

\begin{definition}\label{TC.def}
For an RBES $\mathcal{E}$, 
$\TC(\mathcal{E})$ is a {\em configuration transition system} $(\RConf(\mathcal{E})$, $\rightharpoondown$, $C_0)$,
where $C\stackrel{M}{\rightharpoondown} C'$ in $\TC(\mathcal{E})$
iff
$C\stackrel{(A\cup\underline{B})}{\rightarrow} C'$ in $\mathcal{E}$ and
$M=\sum\limits_{a\in l(A)} Count(a, A)'a\ + \sum\limits_{b\in l(B)} Count(b, B)'\underline{b}=\mathbb{L}(A\cup\underline{B})$,
where the function  $Count(c, C) \colon L \times 2^E \to \mathbb{N}_0$ determines how many events from the set $C$ have the label $c$ as follows:
$Count(c, C) = \mid\{e\in C\mid l(e)=c\}\mid$.
\end{definition}

Let us clarify the definition by providing an illustration.

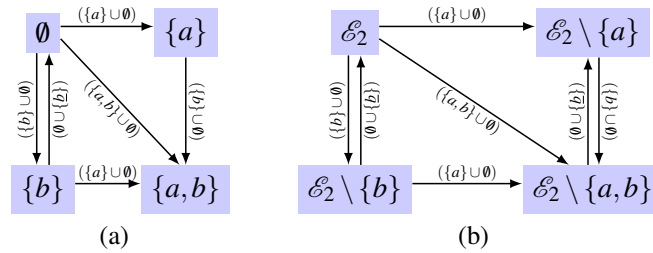
\begin{figure}[htbp]
\centering
\begin{tikzpicture}[line width=0.02cm,>={Latex[length=0.22cm,width=0.1cm]},scale=0.63]
\node[fill=blue!20,at={(1,3)}](b){$\{b\}$};
\node[fill=blue!20,at={(1,6.3)}](E){$\emptyset$};
\node[fill=blue!20,at={(4,3)}](ab){$\{a,b\}$};
\node[fill=blue!20,at={(4,6.3)}](a){$\{a\}$};

\path[-latex]([xshift=0.13 cm]b.north)edge[]node[auto,sloped,below,inner sep=0.03cm,pos=0.5,rotate=0]{\tiny{$(\emptyset \cup\{\underline{b}\})$}}([xshift=0.13 cm]E.south);
\path[-latex]([xshift=-0.13 cm]E.south)edge[]node[auto,sloped,above,inner sep=0.03cm,pos=0.5,rotate=180]{\tiny{$(\{b\}\cup\emptyset)$}}([xshift=-0.13 cm]b.north);

\path[-latex]([yshift=0.13 cm]E.east)edge[]node[auto,sloped,above,inner sep=0.03cm,pos=0.5,rotate=0]{\tiny{$(\{a\}\cup \emptyset)$}}([yshift=0.13cm]a.west);
\path[-latex]([yshift=0.13 cm]b.east)edge[]node[auto,sloped,above,inner sep=0.03cm,pos=0.5,rotate=0]{\tiny{$(\{a\}\cup\emptyset)$}}([yshift=0.13cm]ab.west);

\path[-latex]([yshift=0.0 cm]a.south)edge[]node[auto,sloped,above,inner sep=0.03cm,pos=0.5,rotate=0]{\tiny{$(\{b\}\cup\emptyset)$}}([yshift=0.0cm]ab.north);
\path[-latex]([yshift=-0.13 cm]E.east)edge[]node[auto,sloped,below,inner sep=0.03cm,pos=0.5,rotate=0]{\tiny{$(\{a,b\} \cup \emptyset)$}}([xshift=-0.13cm]ab.north);

\node[at={(2.5,2)}]{\small{(a)}};
\end{tikzpicture}
\hspace*{0.7cm}
\begin{tikzpicture}[line width=0.02cm,>={Latex[length=0.22cm,width=0.1cm]},scale=0.63]
\node[fill=blue!20,at={(1,3)}](b){$\mathcal{E}_2\setminus \{b\}$};
\node[fill=blue!20,at={(1,6.3)}](E){$\mathcal{E}_2$};
\node[fill=blue!20,at={(6,3)}](ab){$\mathcal{E}_2\setminus \{a,b\}$};
\node[fill=blue!20,at={(6,6.3)}](a){$\mathcal{E}_2\setminus \{a\}$};

\path[-latex]([xshift=0.13 cm]b.north)edge[]node[auto,sloped,below,inner sep=0.03cm,pos=0.5,rotate=0]{\tiny{$(\emptyset \cup\{\underline{b}\})$}}([xshift=0.13 cm]E.south);
\path[-latex]([xshift=-0.13 cm]E.south)edge[]node[auto,sloped,above,inner sep=0.03cm,pos=0.5,rotate=180]{\tiny{$(\{b\}\cup\emptyset)$}}([xshift=-0.13 cm]b.north);

\path[-latex]([yshift=0.13 cm]E.east)edge[]node[auto,sloped,above,inner sep=0.03cm,pos=0.5,rotate=0]{\tiny{$(\{a\}\cup\emptyset)$}}([yshift=0.13cm]a.west);
\path[-latex]([yshift=0.13 cm]b.east)edge[]node[auto,sloped,above,inner sep=0.03cm,pos=0.5,rotate=0]{\tiny{$(\{a\}\cup\emptyset)$}}([yshift=0.13cm]ab.west);

\path[-latex]([xshift=-0.1 cm]ab.north)edge[]node[auto,sloped,above,inner sep=0.03cm,pos=0.5,rotate=0]{\tiny{$(\emptyset\cup\{\underline{b}\})$}}([xshift=-0.10cm]a.south);
\path[-latex]([xshift=0.130 cm]a.south)edge[]node[auto,sloped,above,inner sep=0.03cm,pos=0.5,rotate=0]{\tiny{$(\{b\}\cup\emptyset)$}}([xshift=0.130cm]ab.north);
\path[-latex]([yshift=-0.13 cm]E.east)edge[]node[auto,sloped,below,inner sep=0.03cm,pos=0.5,rotate=0]{\tiny{$(\{a,b\}\cup \emptyset)$}}([xshift=-0.5cm]ab.north);
\node[at={(3.5,2)}]{\small{(b)}};
\end{tikzpicture}
\caption{The configuration- and residual-based transition systems for $\mathcal{E}_2$}\label{TC_E2}
\end{figure}

\begin{figure}[htbp]
\centering
\begin{tikzpicture}[line width=0.02cm,>={Latex[length=0.22cm,width=0.1cm]},scale=0.63]
\node[fill=blue!20,at={(1,3.6)}](ac){$\{a,c\}$};
\node[fill=blue!20,at={(1,6.3)}](a){$\{a\}$};
\node[fill=blue!20,at={(6.2,6.3)}](b){$\{b\}$};
\node[fill=blue!20,at={(3.6,6.3)}](E){$\emptyset$};
\node[fill=blue!20,at={(6.2,3.6)}](bc){$\{b,c\}$};
\node[fill=blue!20,at={(1,0.9)}](acd){$\{a,c,d\}$};
\node[fill=blue!20,at={(6.2,0.9)}](bcd){$\{b,c,d\}$};

\path[-latex]([xshift=0.13 cm]ac.north)edge[]node[auto,sloped,below,inner sep=0.03cm,pos=0.5,rotate=0]{\tiny{$(\emptyset \cup\{\underline{c}\})$}}([xshift=0.13 cm]a.south);
\path[-latex]([xshift=-0.13 cm]a.south)edge[]node[auto,sloped,above,inner sep=0.03cm,pos=0.5,rotate=180]{\tiny{$(\{c\}\cup\emptyset)$}}([xshift=-0.13 cm]ac.north);

\path[-latex]([yshift=0.13 cm]a.east)edge[]node[auto,sloped,above,inner sep=0.03cm,pos=0.5,rotate=0]{\tiny{$(\emptyset\cup\{\underline{a}\})$}}([yshift=0.13cm]E.west);
\path[-latex]([yshift=-0.13 cm]E.west)edge[]node[auto,sloped,below,inner sep=0.03cm,pos=0.5,rotate=0]{\tiny{$(\{a\}\cup\emptyset)$}}([yshift=-0.13cm]a.east);

\path[-latex]([yshift=0.0 cm]E.east)edge[]node[auto,sloped,above,inner sep=0.03cm,pos=0.5,rotate=0]{\tiny{$(\{b\}\cup\emptyset)$}}([yshift=0.0cm]b.west);

\path[-latex]([xshift=0.13 cm]bc.north)edge[]node[auto,sloped,below,inner sep=0.03cm,pos=0.5,rotate=0]{\tiny{$(\emptyset \cup\{\underline{c}\})$}}([xshift=0.13 cm]b.south);
\path[-latex]([xshift=-0.13 cm]b.south)edge[]node[auto,sloped,above,inner sep=0.03cm,pos=0.5,rotate=180]{\tiny{$(\{c\}\cup\emptyset)$}}([xshift=-0.13 cm]bc.north);

\path[-latex]([xshift=0.0 cm]ac.south)edge[]node[auto,sloped,above,inner sep=0.03cm,pos=0.5,rotate=180]{\tiny{$(\{d\}\cup\emptyset)$}}([xshift=0.0 cm]acd.north);

\path[-latex]([xshift=0.0 cm]bc.south)edge[]node[auto,sloped,above,inner sep=0.03cm,pos=0.5,rotate=180]{\tiny{$(\{d\}\cup\emptyset)$}}([xshift=0.0 cm]bcd.north);
\node[at={(3.5,0.)}]{\small{(a)}};
\end{tikzpicture}
\hspace*{0.4cm}
\begin{tikzpicture}[line width=0.02cm,>={Latex[length=0.22cm,width=0.1cm]},scale=0.63]
\node[fill=blue!20,at={(1,3.6)}](ac){$\mathcal{E}_3\setminus\{a,c\}$};
\node[fill=blue!20,at={(1,6.3)}](a){$\mathcal{E}_3\setminus\{a\}$};
\node[fill=blue!20,at={(8,6.3)}](b){$\mathcal{E}_3\setminus\{b\}$};
\node[fill=blue!20,at={(4.5,6.3)}](E){$\mathcal{E}_3\setminus\emptyset$};
\node[fill=blue!20,at={(8,3.6)}](bc){$\mathcal{E}_3\setminus\{b,c\}$};
\node[fill=blue!20,at={(1,0.9)}](acd){$\mathcal{E}_3\setminus\{a,c,d\}$};
\node[fill=blue!20,at={(8,0.9)}](bcd){$\mathcal{E}_3\setminus\{b,c,d\}$};

\path[-latex]([xshift=0.13 cm]ac.north)edge[]node[auto,sloped,below,inner sep=0.03cm,pos=0.5,rotate=0]{\tiny{$(\emptyset \cup\{\underline{c}\})$}}([xshift=0.13 cm]a.south);
\path[-latex]([xshift=-0.13 cm]a.south)edge[]node[auto,sloped,above,inner sep=0.03cm,pos=0.5,rotate=180]{\tiny{$(\{c\}\cup\emptyset)$}}([xshift=-0.13 cm]ac.north);

\path[-latex]([yshift=0.13 cm]a.east)edge[]node[auto,sloped,above,inner sep=0.03cm,pos=0.5,rotate=0]{\tiny{$(\emptyset\cup\{\underline{a}\})$}}([yshift=0.13cm]E.west);
\path[-latex]([yshift=-0.13 cm]E.west)edge[]node[auto,sloped,below,inner sep=0.03cm,pos=0.5,rotate=0]{\tiny{$(\{a\}\cup\emptyset)$}}([yshift=-0.13cm]a.east);

\path[-latex]([yshift=0.0 cm]E.east)edge[]node[auto,sloped,above,inner sep=0.03cm,pos=0.5,rotate=0]{\tiny{$(\{b\}\cup\emptyset)$}}([yshift=0.0cm]b.west);

\path[-latex]([xshift=0.13 cm]bc.north)edge[]node[auto,sloped,below,inner sep=0.03cm,pos=0.5,rotate=0]{\tiny{$(\emptyset \cup\{\underline{c}\})$}}([xshift=0.13 cm]b.south);
\path[-latex]([xshift=-0.13 cm]b.south)edge[]node[auto,sloped,above,inner sep=0.03cm,pos=0.5,rotate=180]{\tiny{$(\{c\}\cup\emptyset)$}}([xshift=-0.13 cm]bc.north);

\path[-latex]([xshift=0.0 cm]ac.south)edge[]node[auto,sloped,above,inner sep=0.03cm,pos=0.5,rotate=180]{\tiny{$(\{d\}\cup\emptyset)$}}([xshift=0.0 cm]acd.north);

\path[-latex]([xshift=0.0 cm]bc.south)edge[]node[auto,sloped,above,inner sep=0.03cm,pos=0.5,rotate=180]{\tiny{$(\{d\}\cup\emptyset)$}}([xshift=0.0 cm]bcd.north);
\node[at={(4.5,0.)}]{\small{(b)}};
\end{tikzpicture}
\caption{The configuration- and residual-based transition systems for $\mathcal{E}_3$}\label{TC_E3}
\end{figure}
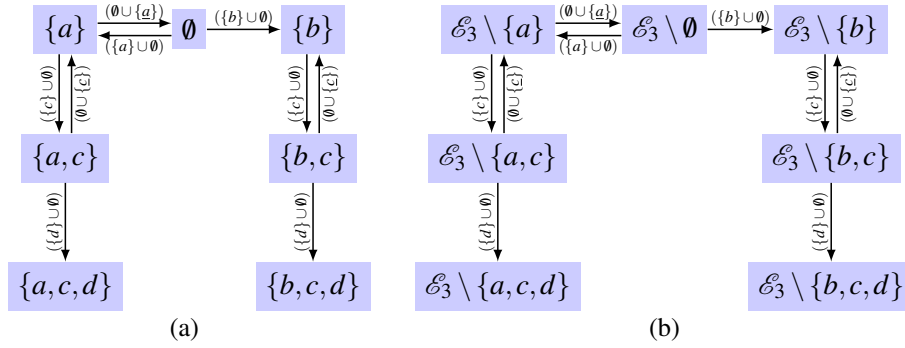

\begin{example}\label{8.examp}
First, consider the cause-respecting RBES $\mathcal{E}_2$ which is shown in Fig.~\ref{R_Bund}(b) and discussed in Examples~\ref{3.examp}--\ref{4.examp}.
Recall that $RConf(\mathcal{E}_2) = \{\emptyset$, $\{a\}$, $\{b\}$, $\{a,b\}\}$.
Applying Definition~\ref{TC.def}, we obtain the configuration transition system $\TC(\mathcal{E}_2)$, which is drawn in Fig.~\ref{TC_E2}(a).

Second, contemplate the causal RBES $\mathcal{E}_3$ which is presented in Fig.~\ref{R_Bund}(c) and examined in Examples~\ref{3.examp}--\ref{4.examp}.
We know that $RConf(\mathcal{E}_3) = \{\{a\}, \emptyset, \{b\}, \{a,c\}, \{b,c\}, \{a,c,d\}, \{b,c,d\}\}$.
Using Definition~\ref{TC.def}, we construct the configuration transition system $\TC(\mathcal{E}_3)$, which is depicted in Fig.~\ref{TC_E3}(a).
\hfill$\Diamond$
\end{example}

Introduce auxiliary notations.
For RBESs
$\mathcal{E}$ and $\mathcal{E}'$,
we shall write
$\mathcal{E}\stackrel{(A\cup\underline{B})}{\rightarrow}\mathcal{E}'$ iff
$C_0\stackrel{(A\cup\underline{B})}{\rightarrow} C$ in $\mathcal{E}$ and $\mathcal{E}'=\mathcal{E}\setminus C$.
Let $Reach(\mathcal{E})=\{\mathcal{F} \mid\exists\mathcal{E}_0, \ldots, \mathcal{E}_k$ $(k\geq 0)$ s.t.
$\mathcal{E}_0 = \mathcal{E}$, $\mathcal{E}_k=\mathcal{F}$, and
$\mathcal{E}_{i-1}\stackrel{(A\cup\underline{B})}{\rightarrow}\mathcal{E}_{i}$ $(1\leq i\leq k)\}$.

We next propose the definition of labeled transition systems with RBESs as states.

\begin{definition}\label{TR.def}
For an RBES $\mathcal{E}$,
$\TE(\mathcal{E})$ is a {\em residual transition system} $(Reach(\mathcal{E})$, $\stackrel{}{\rightharpoonup}$, $\mathcal{E})$,
where $\mathcal{F}\stackrel{M}{\rightharpoonup}\mathcal{F}'$ in $\TE(\mathcal{E})$ iff
$\mathcal{F}\stackrel{(A\cup\underline{B})}{\rightarrow}\mathcal{F}'$
and
$M=\sum\limits_{a\in l(A)} Count(a, A)'a\ + \sum\limits_{b\in l(B)} Count(b, B)'\underline{b}=\mathbb{L}(A\cup\underline{B})$.
\end{definition}

We give an example illustrating the above definition.

\begin{example}\label{9.examp}
The residuals of the cause-respecting RBES $\mathcal{E}_2$ are drawn in Fig.~\ref{rem_RBES2}.
Using Definition~\ref{TR.def}, we form the residual transition system $\TR(\mathcal{E}_2)$, which is demonstrated in Fig.~\ref{TC_E2}(b).

The residuals of the causal RBES $\mathcal{E}_3$ are discussed in Example~\ref{5.examp} and shown in Fig.~\ref{rem_RBES3}.
Applying Definition~\ref{TR.def}, we construct the residual transition system $\TR(\mathcal{E}_3)$, which is depicted in Fig.~\ref{TC_E3}(b).
\hfill$\Diamond$
\end{example}

We establish the relationships between states and transitions of the two types of transition systems for the causal RBES.

\begin{proposition}\label{prop4}
Given a causal RBES $\mathcal{E}=(E, F, \sharp, \mapsto, \rhd, l, C_0)$, it holds:
\begin{itemize}
\item[(i)] for any $C \in RConf(\mathcal{E})$, $\mathcal{E}\setminus C\in Reach(\mathcal{E})$;

\item[(ii)] for any $\mathcal{E}'\in Reach(\mathcal{E})$, there is a unique $C \in RConf(\mathcal{E})$ such that $\mathcal{E}'=\mathcal{E}\setminus C$;

\item[(iii)]
if $C\stackrel{M}{\rightharpoondown} C'$ in $TC(\mathcal{E})$
then $\mathcal{E}\setminus C \stackrel{M}{\rightharpoonup} \mathcal{E} \setminus C'$ in $\TE(\mathcal{E})$;

\item[(iv)]
if $\mathcal{E}'\stackrel{M}\rightharpoonup\mathcal{E}''$ in $\TE(\mathcal{E})$, then
there exist $C', C''\in RConf(\mathcal{E})$ such that $\mathcal{E}'=\mathcal{E}\setminus C'$, $\mathcal{E}''=\mathcal{E}\setminus C''$, and
$C'\stackrel{M}{\rightharpoondown} C''$ in $TC(\mathcal{E})$.
\end{itemize}
\end{proposition}

We state the main result of the paper.
\begin{theorem}\label{main.th}
Given a causal RBES $\mathcal{E}$, $\TC(\mathcal{E})$ and $\TR(\mathcal{E})$ are isomorphic.
\end{theorem}

\begin{example}\label{10.examp}
It is obvious that the configuration- and residual-based transition systems for the causal RBES $\mathcal{E}_3$, which are depicted in Fig.~\ref{TC_E3}, are isomorphic.
On the other hand, the cause-respecting and non-causal RBES $\mathcal{E}_2$ has the non-isomorphic configuration- and residual-based transition systems, which are presented in Fig.~\ref{TC_E2}.
\hfill$\Diamond$
\end{example}

\section{Considering $TC(\cdot)$ and $TR(\cdot)$ as Functors}
In this section, we first consider the definitions of categories of LTSs and RBESs,
and then determine whether the mappings $\TC(\mathcal{E})$ and $\TR(\mathcal{E})$, where $\mathcal{E}$ is a causal RBES, are the functors between the categories.

We recall the concept of an LTS-morphism in Definition~\ref{def_morphism_1},
giving us the category $\mathbf{LTS}_{\mathbb{L}}$ with LTSs labeled over $\mathbb{L}$ as objects.

\begin{definition}\label{def_morphism_1}
Given transition systems $\mathcal{T} = (S$, $\rightarrow$, $i)$ and $\mathcal{T}' = (S'$, $\rightarrow'$, $i')$ labeled over $\mathbb{L}$,
a mapping $\nu : S \to S'$ is a {\em morphism} from $\mathcal{T}$ to $\mathcal{T}'$, if:
$\nu(i)=i'$, and for all $s,s_1\in S$ and for all $\lambda\in\mathbb{L}$ it holds:
if $(s,\lambda,s_1)\in\rightarrow$ in $\mathcal{T}$ then $(\nu(s),\lambda,\nu(s_1))\in\rightarrow'$ in $\mathcal{T}'$.
\end{definition}

Observe that the morphism between two LTSs is a simulation of each other's transitions.

\begin{example}\label{11.examp}
Consider the transition systems $TC(\mathcal{E}_2)$ and $TR(\mathcal{E}_2)$ from Example~\ref{9.examp} (see Fig.~\ref{TC_E2}).
Define a mapping $\nu : Conf(\mathcal{E}_2) \to Reach(\mathcal{E}_2)$ as follows: $\nu(C) = \mathcal{E}_2 \setminus C$ for all $C\in Conf(\mathcal{E}_2)$.
It is clear that $\nu(C^2_0) = \mathcal{E}_2\setminus C^2_0 = \mathcal{E}_2$, thanks to Lemma~\ref{lem_rem_0}(ii).
Then, it is easy to see that $\nu$ is indeed a morphism from $TC(\mathcal{E}_2)$ to $TR(\mathcal{E}_2)$, using Definition~\ref{def_morphism_1}.
  \hfill$\Diamond$
\end{example}

We next consider a slight modification of the definition of the RBES-morphism from \cite{GPY21},
which allows us to deal with the category $\mathbf{RBES}_{L}$, whose objects are RBESs with labels from $L$.
In addition, to define the RBES-morphisms,
we use total rather than partial mappings in order to ensure consistency with the LTS-morphisms, which are total mappings.

\begin{definition}\label{def_morphism}
Given RBESs
$\mathcal{E} = (E$, $F$, $\sharp$, $\mapsto$, $\rhd$, $l$, $C_0)$ and $\mathcal{E}' = (E'$, $F'$, $\sharp'$, $\mapsto'$, $\rhd'$, $l'$, $C'_0)$,
a mapping $\mu : E \to E'$ is a {\em morphism} from $\mathcal{E}$ to $\mathcal{E}'$, if:
$\mu(F)\subseteq F'$, $\mu(C_0)=C'_0$, and for all $e,e'\in E$ it holds:
\begin{itemize}
\item[(a)] if $\mu(e)\ \sharp'\ \mu(e')$ then $e\ \sharp\ e'$;
\item[(b)] if $\mu(e)=\mu(e')$ and $e\neq e'$, then $e\ \sharp\ e'$;
\item[(c)] for $W'\subseteq E'$ if $W'\mapsto'\mu(e)^\ast$ then there is $W\subseteq E$ such that $W\mapsto e^\ast$ and $\mu(W)\subseteq W'$; 
\item[(d)] if $\mu(e) \rhd'\underline{\mu(e')}$ then $e\rhd \underline{e'}$;
\item[(e)] $l' \circ \mu = l$.
\end{itemize}
\end{definition}

\begin{example}\label{12.examp}
Consider the RBES $\mathcal{E}_2$ from Example~\ref{3.examp} (see Fig.~\ref{R_Bund}(b)) and the the RBES $\mathcal{E}_4$ with the components:
$E_4=\{a,b,c\}$; $F_4=\{b,c\}$;
$\sharp_4 = \emptyset$; $\mapsto_4 =  \{(\{b\}, \underline{b}), (\{c\}, \underline{c})\}$;
$\rhd_4 = \emptyset$;
$C^4_0 = \emptyset$.
Define a mapping $\mu : E_2 \to E_4$ as follows: $\mu(a)=a$ and $\mu(b)=b$. 
It is clear that
$\mu(F_2) = \{b\}\subseteq F_4$,   $\mu(C^2_0=\emptyset)=\emptyset = C^4_0$, and $l_4 \circ \mu = l_2$.
Check that the items of Definition~\ref{def_morphism} hold.
Item (a) is true because $\sharp_4=\emptyset$.
Item (b) is correct since $\mu$ is an injective mapping. 
As $\mu(E_2) = \{a,b\}$ and $\mapsto_4 =  \{(\{b\}, \underline{b}), (\{c\}, \underline{c})\}$,
we obtain $W'=\{b\} \mapsto_4\mu(e)^\ast = \underline{\mu(b)} = \underline{b}$.
Then, we have that $W=\{b\} \mapsto_2 \underline{b}$ and $\mu(\{b\}) = \{b\} = W'$.
So, item (c) holds.
Item (d) follows from the fact $\rhd_4=\emptyset$.
Thus, $\mu$ is indeed a morphism from $\mathcal{E}_2$ to $\mathcal{E}_4$, by Definition~\ref{def_morphism}.
\hfill$\Diamond$
\end{example}

We demonstrate that the RBES-morphism represents the idea of simulating the behavior of one RBES by the behavior of another RBES.
\begin{lemma}\label{lemma.morph}
Given objects $\mathcal{E}$, $\mathcal{E}'$ and a morphism $\mu : \mathcal{E} \to \mathcal{E}'$ in the category $\mathbf{RBES}_{L}$,
$\mu(C)\in RConf(\mathcal{E}')$ for all $C\in RConf(\mathcal{E})$,
and whenever $C\xrightarrow{(A\cup\underline{B})} C_1$ in $\mathcal{E}$,
then $\mu(C) 
\xrightarrow{(\mu(A)\cup\underline{\mu(B)})}
\mu(C_1)$ in $\mathcal{E}'$.
\end{lemma}

In order to extend the mapping $\TC(\cdot)$ ($\TR(\cdot)$) to a functor from the category $\mathbf{RBES}_{L}$ to the category $\mathbf{LTS}_{\mathbb{L}}$,
define how for objects RBESs $\mathcal{E}$ and $\mathcal{E}'$ in $\mathbf{RBES}_{L}$ this mapping transforms morphisms in $\mathbf{RBES}_{L}$
into morphisms in $\mathbf{LTS}_{\mathbb{L}}$:
for a morphism $\mu\colon E\rightarrow E'$ in $\mathbf{RBES}_{L}$,
set $TC(\mu)(C)=\mu(C)$ for any $C\in RConf(\mathcal{E})$
($TR(\mu)(\mathcal{F})=\mathcal{E}'\setminus\mu(C)$ for any $\mathcal{F}\in Reach(\mathcal{E})$
such that $\mathcal{F} = \mathcal{E} \setminus C$ for some $C\in RConf(\mathcal{E})$).

\begin{proposition}\label{prop.funct}
Given a causal RBES,
the mapping $TC(\mathcal{E})$ ($TR(\mathcal{E})$) yields a functor from the category $\mathbf{RBES}_{L}$ to the category $\mathbf{LTS}_{\mathbb{L}}$.
\end{proposition}

\section{Concluding Remarks}

In this paper,
we dealt with two different --  configuration- and residual-based -- ways of giving (step) transition system semantics for reversible bundle event structures (RBESs).
For this purpose, we firstly defined (step) semantics from \cite{GPY21,GPY22},
which is based on configurations obtained when starting from an initial configuration and executing concurrent events and/or undoing previously executed events,
and we secondly developed a removal operator which is useful for constructing residuals (model fragments) by retaining an appropriate amount of structure remaining during the execution of the model.
In addition, we stated some correctness criteria for the removal operator.
The meaning of the correctness properties is that the obtained residuals do not allow configurations that are disallowed by the original structure.
As our main result, we obtained an isomorphism between configuration- and residual-based transition systems from the causal RBES.
In addition, we have shown that the mappings $\TC(\mathcal{E})$ and $\TE(\mathcal{E})$, where $\mathcal{E}$ is a causal RBES,
are functors from the category $\mathbf{RBES}_{L}$ to the category $\mathbf{LTS}_{\mathbb{L}}$.

We expect that, due to our results concerning isomorphisms,
many interesting facts known from the literature on configuration transition systems (e.g., see \cite{BCG17,GP09,W89} among others) can be extended to residual transition systems.
Another benefit of having isomorphism instead of bisimulation arises in the context of (reversible) prime event structures (a subclass of (R)BESs).
With bisimulation between $\TC(\cdot)$ and $\TE(\cdot)$,
the papers \cite{MR98,GV25} have shown that the operator $\TC$ is a functor from a category of classical/reversible prime event structures
into a category of transition systems, in interleaving and step semantics, whereas the operator $\TE$ is not in either of these semantics.
This is at variance with a postulate by Winskel and Nielsen \cite{WN95} that any semantic model should form a category and its semantic operations should possess a categorical characterization.
Isomorphisms between the two kinds of transition systems conform better to this postulate.
Also, we anticipate that the isomorphisms obtained in this paper will make it possible to closely relate transition systems constructed on configurations and
transition systems derived from denotational event structure semantics of reversible process calculi.
In \cite{GPY22}, the authors have defined $\pi IH$ and $\pi IK$, the first reversible early $\pi$-calculus.
In $\pi IH$, reversible actions are tracked by using extrusion histories, while moving reversible actions and their locations into separate histories to ensure dynamic reversibility.
In contrast, static $\pi IK$ maintains the structure of the process intact while annotating reversible actions with keys.
The denotational semantics of $\pi IK$ in terms of RBESs is constructed inductively on the syntax of the $\pi IK$-process, and
a correspondence between a $\pi IK$-process and the causal RBES it generates has been demonstrated.
The operational semantics of $\pi IH$, which generates a reversible prime event structure (RPES) based on a labelled asynchronous transition system, has been defined.
A correspondence between the resulting RPESs and RBESs has been established.
Recall that in several papers (see \cite{BM94,BC89,CVY12,K96,L93} among others),
residual-based transition systems are actively used in providing operational semantics of different irreversible process calculi and
in demonstrating the consistency of operational and denotational semantics.
So, we believe that it would be appropriate to develop operational semantics of the above reversible process calculi in terms of residual-based transition systems from RBESs.

Thus, the advantage of the approach proposed in the paper is that
it achieves isomorphism between two different types of transition systems from a single RBES using non-executable events.
The disadvantage is that the removal operator reduces the number of events in RBES only by deleting those
that have already been executed, leaving conflicting events (which cannot be executed in the future) as non-executable.

As for future work, we have a plan to study properties of the functors presented in Section~5.
Also, we intend to investigate cases where the removal operator on RBESs developed here matches the residuation operation on configuration structures
proposed in the paper \cite{AK25}.
Work is underway to extend our approach to non-causal reversible event structures,
which yields promising intermediate results but requires a more sophisticated removal operator.
Another line of future research is to generalize the model of reversible prime event structures with non-executable events
(for example, by dropping the transitivity/acyclicity of causality, as well as the principles of finite causes).
We expect that this will allow us to obtain isomorphisms between the two types of transition systems from the models, as was done for extended prime event structures in the paper \cite{BGV18}.
In addition, we plan to broaden the list of studied models with reversible versions of flow/\-stable/\-general event structures with symmetric/asymmetric conflict.

\bibliographystyle{eptcs}
\bibliography{genpapers.bib}

\begin{thebibliography}{10}
\providecommand{\bibitemdeclare}[2]{}
\providecommand{\surnamestart}{}
\providecommand{\surnameend}{}
\providecommand{\urlprefix}{Available at }
\providecommand{\url}[1]{\texttt{#1}}
\providecommand{\href}[2]{\texttt{#2}}
\providecommand{\urlalt}[2]{\href{#1}{#2}}
\providecommand{\doi}[1]{doi:\urlalt{https://doi.org/#1}{#1}}
\providecommand{\eprint}[1]{arXiv:\urlalt{https://arxiv.org/abs/#1}{#1}}
\providecommand{\bibinfo}[2]{#2}

\bibitemdeclare{inproceedings}{AK25}
\bibitem{AK25}
\bibinfo{author}{Cl{\'{e}}ment \surnamestart Aubert\surnameend} \&
  \bibinfo{author}{Jean \surnamestart Krivine\surnameend}
  (\bibinfo{year}{2025}): \emph{\bibinfo{title}{Reversible computations are
  computations}}.
\newblock In \bibinfo{editor}{Clemens \surnamestart Kupke\surnameend} \&
  \bibinfo{editor}{Stefan \surnamestart Milius\surnameend}, editors: {\slshape
  \bibinfo{booktitle}{Proceedings of the 41st Conference on the Mathematical
  Foundations of Programming Semantics, {MFPS} XLI, University of Strathclyde,
  Glasgow, UK, June 16-21, 2025}}, {\slshape \bibinfo{series}{Electronic Notes
  in Theoretical Informatics and Computer Science}}~\bibinfo{volume}{5},
  \bibinfo{publisher}{EpiSciences}, \doi{10.46298/ENTICS.16677}.

\bibitemdeclare{article}{BM94}
\bibitem{BM94}
\bibinfo{author}{Christel \surnamestart Baier\surnameend} \&
  \bibinfo{author}{Mila \surnamestart Majster-Cederbaum\surnameend}
  (\bibinfo{year}{1994}): \emph{\bibinfo{title}{The Connection between an Event
  Structure Semantics and an Operational Semantics for TCSP}}.
\newblock {\slshape \bibinfo{journal}{Acta Informatica}}
  \bibinfo{volume}{31}(\bibinfo{number}{1}), \doi{10.1007/BF01178923}.

\bibitemdeclare{inproceedings}{BCG17}
\bibitem{BCG17}
\bibinfo{author}{Paolo \surnamestart Baldan\surnameend},
  \bibinfo{author}{Andrea \surnamestart Corradini\surnameend} \&
  \bibinfo{author}{Fabio \surnamestart Gadducci\surnameend}
  (\bibinfo{year}{2017}): \emph{\bibinfo{title}{Domains and Event Structures
  for Fusions}}.
\newblock In: {\slshape \bibinfo{booktitle}{32nd Annual {ACM/IEEE} Symposium on
  Logic in Computer Science, {LICS} 2017, Reykjavik, Iceland, June 20-23,
  2017}}, \bibinfo{publisher}{{IEEE} Computer Society}, pp.
  \bibinfo{pages}{1--12}, \doi{10.1109/LICS.2017.8005135}.

\bibitemdeclare{inproceedings}{BGV17}
\bibitem{BGV17}
\bibinfo{author}{Eike \surnamestart Best\surnameend}, \bibinfo{author}{Nataliya
  \surnamestart Gribovskaya\surnameend} \& \bibinfo{author}{Irina \surnamestart
  Virbitskaite\surnameend} (\bibinfo{year}{2017}):
  \emph{\bibinfo{title}{Configuration- and Residual-Based Transition Systems
  for Event Structures with Asymmetric Conflict}}.
\newblock In \bibinfo{editor}{Bernhard \surnamestart Steffen\surnameend},
  \bibinfo{editor}{Christel \surnamestart Baier\surnameend},
  \bibinfo{editor}{Mark \surnamestart van~den Brand\surnameend},
  \bibinfo{editor}{Johann \surnamestart Eder\surnameend}, \bibinfo{editor}{Mike
  \surnamestart Hinchey\surnameend} \& \bibinfo{editor}{Tiziana \surnamestart
  Margaria\surnameend}, editors: {\slshape \bibinfo{booktitle}{SOFSEM 2017:
  Theory and Practice of Computer Science}}, \bibinfo{publisher}{Springer
  International Publishing}, \bibinfo{address}{Cham}, pp.
  \bibinfo{pages}{132--146}, \doi{10.1007/978-3-319-51963-0_11}.

\bibitemdeclare{inproceedings}{BGV18}
\bibitem{BGV18}
\bibinfo{author}{Eike \surnamestart Best\surnameend}, \bibinfo{author}{Nataliya
  \surnamestart Gribovskaya\surnameend} \& \bibinfo{author}{Irina \surnamestart
  Virbitskaite\surnameend} (\bibinfo{year}{2018}): \emph{\bibinfo{title}{From
  Event-Oriented Models to Transition Systems}}.
\newblock In \bibinfo{editor}{Victor \surnamestart Khomenko\surnameend} \&
  \bibinfo{editor}{Olivier~H. \surnamestart Roux\surnameend}, editors:
  {\slshape \bibinfo{booktitle}{Application and Theory of Petri Nets and
  Concurrency}}, \bibinfo{publisher}{Springer International Publishing},
  \bibinfo{address}{Cham}, pp. \bibinfo{pages}{117--139},
  \doi{10.1007/978-3-319-91268-4_7}.

\bibitemdeclare{article}{BC89}
\bibitem{BC89}
\bibinfo{author}{Gerard \surnamestart Boudol\surnameend} \&
  \bibinfo{author}{Ilaria \surnamestart Castellani\surnameend}
  (\bibinfo{year}{1988}): \emph{\bibinfo{title}{Concurrency and Atomicity}}.
\newblock {\slshape \bibinfo{journal}{Theoretical Computer Science}}
  \bibinfo{volume}{59}(\bibinfo{number}{1}), pp. \bibinfo{pages}{25--84},
  \doi{10.1016/0304-3975(88)90096-5}.

\bibitemdeclare{inproceedings}{CVY12}
\bibitem{CVY12}
\bibinfo{author}{Silvia \surnamestart Crafa\surnameend},
  \bibinfo{author}{Daniele \surnamestart Varacca\surnameend} \&
  \bibinfo{author}{Nobuko \surnamestart Yoshida\surnameend}
  (\bibinfo{year}{2012}): \emph{\bibinfo{title}{Event Structure Semantics of
  Parallel Extrusion in the $\pi$-Calculus}}.
\newblock In \bibinfo{editor}{Lars \surnamestart Birkedal\surnameend}, editor:
  {\slshape \bibinfo{booktitle}{Foundations of Software Science and
  Computational Structures}}, \bibinfo{publisher}{Springer Berlin Heidelberg},
  \bibinfo{address}{Berlin, Heidelberg}, pp. \bibinfo{pages}{225--239},
  \doi{10.1007/978-3-642-28729-9_15}.

\bibitemdeclare{inproceedings}{GLMMPUV23}
\bibitem{GLMMPUV23}
\bibinfo{author}{Robert \surnamestart Gl{\"{u}}ck\surnameend},
  \bibinfo{author}{Ivan \surnamestart Lanese\surnameend},
  \bibinfo{author}{Claudio~Antares \surnamestart Mezzina\surnameend},
  \bibinfo{author}{Jaroslaw~Adam \surnamestart Miszczak\surnameend},
  \bibinfo{author}{Iain \surnamestart Phillips\surnameend},
  \bibinfo{author}{Irek \surnamestart Ulidowski\surnameend} \&
  \bibinfo{author}{Germ{\'{a}}n \surnamestart Vidal\surnameend}
  (\bibinfo{year}{2023}): \emph{\bibinfo{title}{Towards a Taxonomy for
  Reversible Computation Approaches}}.
\newblock In \bibinfo{editor}{Martin \surnamestart Kutrib\surnameend} \&
  \bibinfo{editor}{Uwe \surnamestart Meyer\surnameend}, editors: {\slshape
  \bibinfo{booktitle}{Reversible Computation - 15th International Conference,
  {RC} 2023, Giessen, Germany, July 18-19, 2023, Proceedings}}, {\slshape
  \bibinfo{series}{Lecture Notes in Computer Science}} \bibinfo{volume}{13960},
  \bibinfo{publisher}{Springer}, pp. \bibinfo{pages}{24--39},
  \doi{10.1007/978-3-031-38100-3\_3}.

\bibitemdeclare{article}{GPY21}
\bibitem{GPY21}
\bibinfo{author}{Eva \surnamestart Graversen\surnameend}, \bibinfo{author}{Iain
  \surnamestart Phillips\surnameend} \& \bibinfo{author}{Nobuko \surnamestart
  Yoshida\surnameend} (\bibinfo{year}{2021}): \emph{\bibinfo{title}{Event
  Structure Semantics of (Controlled) Reversible CCS}}.
\newblock {\slshape \bibinfo{journal}{Journal of Logical and Algebraic Methods
  in Programming}} \bibinfo{volume}{121}, p. \bibinfo{pages}{100686},
  \doi{10.1016/j.jlamp.2021.100686}.

\bibitemdeclare{misc}{GPY22}
\bibitem{GPY22}
\bibinfo{author}{Eva \surnamestart Graversen\surnameend}, \bibinfo{author}{Iain
  \surnamestart Phillips\surnameend} \& \bibinfo{author}{Nobuko \surnamestart
  Yoshida\surnameend} (\bibinfo{year}{2022}): \emph{\bibinfo{title}{Event
  Structures for the Reversible Early Internal pi-Calculus}},
  \doi{10.1016/j.jlamp.2021.100720}.

\bibitemdeclare{article}{GV25}
\bibitem{GV25}
\bibinfo{author}{Nataliya \surnamestart Gribovskaya\surnameend} \&
  \bibinfo{author}{Irina \surnamestart Virbitskaite\surnameend}
  (\bibinfo{year}{2025}): \emph{\bibinfo{title}{Transition Systems from
  Asymmetric Prime Event Structures with Cause-Respecting Reversibility}}.
\newblock {\slshape \bibinfo{journal}{International Journal of Foundations of
  Computer Science}} \bibinfo{volume}{0}(\bibinfo{number}{0}), pp.
  \bibinfo{pages}{1--29}, \doi{10.1142/S0129054125460049}.

\bibitemdeclare{inproceedings}{GV23}
\bibitem{GV23}
\bibinfo{author}{Nataliya \surnamestart Gribovskaya\surnameend} \&
  \bibinfo{author}{Irina~B. \surnamestart Virbitskaite\surnameend}
  (\bibinfo{year}{2023}): \emph{\bibinfo{title}{Comparative Transition System
  Semantics for Cause-Respecting Reversible Prime Event Structures}}.
\newblock In \bibinfo{editor}{Zsolt \surnamestart Gazdag\surnameend},
  \bibinfo{editor}{Szabolcs \surnamestart Iv{\'{a}}n\surnameend} \&
  \bibinfo{editor}{Gergely \surnamestart Kov{\'{a}}sznai\surnameend}, editors:
  {\slshape \bibinfo{booktitle}{Proceedings of the 16th International
  Conference on Automata and Formal Languages, {AFL} 2023, Eger, Hungary,
  September 5-7, 2023}}, {\slshape \bibinfo{series}{{EPTCS}}}
  \bibinfo{volume}{386}, pp. \bibinfo{pages}{112--126},
  \doi{10.4204/EPTCS.386.10}.

\bibitemdeclare{phdthesis}{K96}
\bibitem{K96}
\bibinfo{author}{Joost-Pieter \surnamestart Katoen\surnameend}
  (\bibinfo{year}{1996}): \emph{\bibinfo{title}{Quantitative and Qualitative
  Extensions of Event Structures}}.
\newblock Ph.D. thesis, \bibinfo{school}{University of Twente},
  \bibinfo{address}{Netherlands}.

\bibitemdeclare{inproceedings}{L93}
\bibitem{L93}
\bibinfo{author}{Rom \surnamestart Langerak\surnameend} (\bibinfo{year}{1991}):
  \emph{\bibinfo{title}{Bundle Event Structures: a Non-interleaving Semantics
  for LOTOS}}.
\newblock In \bibinfo{editor}{Michel \surnamestart Diaz\surnameend} \&
  \bibinfo{editor}{Roland \surnamestart Groz\surnameend}, editors: {\slshape
  \bibinfo{booktitle}{Formal Description Techniques V}}, \bibinfo{series}{IFIP
  transactions C, Communication systems}, \bibinfo{publisher}{North Holland},
  \bibinfo{address}{Netherlands}, pp. \bibinfo{pages}{331--346}.
\newblock \bibinfo{note}{5th International Conference on Formal Description
  Techniques for Distributed Systems and Communications Protocols, FORTE 1992}.

\bibitemdeclare{article}{MR98}
\bibitem{MR98}
\bibinfo{author}{Mila \surnamestart Majster-Cederbaum\surnameend} \&
  \bibinfo{author}{Markus \surnamestart Roggenbach\surnameend}
  (\bibinfo{year}{1998}): \emph{\bibinfo{title}{Transition Systems from Event
  Structures Revisited}}.
\newblock {\slshape \bibinfo{journal}{Information Processing Letters}}
  \bibinfo{volume}{67}(\bibinfo{number}{3}), pp. \bibinfo{pages}{119--124},
  \doi{10.1016/S0020-0190(98)00105-7}.

\bibitemdeclare{inbook}{WN95}
\bibitem{WN95}
\bibinfo{author}{Mogens \surnamestart Nielsen\surnameend} \&
  \bibinfo{author}{Glynn \surnamestart Winskel\surnameend}
  (\bibinfo{year}{1995}): \emph{\bibinfo{title}{Models for Concurrency}}, pp.
  \bibinfo{pages}{1--148}.
\newblock \bibinfo{publisher}{Oxford University Press}.
\newblock \bibinfo{note}{Also published in BRICS Research Series as RS-94-12}.

\bibitemdeclare{article}{PU15}
\bibitem{PU15}
\bibinfo{author}{Iain \surnamestart Phillips\surnameend} \&
  \bibinfo{author}{Irek \surnamestart Ulidowski\surnameend}
  (\bibinfo{year}{2015}): \emph{\bibinfo{title}{Reversibility and Asymmetric
  Conflict in Event Structures}}.
\newblock {\slshape \bibinfo{journal}{Journal of Logical and Algebraic Methods
  in Programming}} \bibinfo{volume}{84}(\bibinfo{number}{6}), pp.
  \bibinfo{pages}{781--805}, \doi{10.1016/j.jlamp.2015.07.004}.

\bibitemdeclare{article}{UPY18}
\bibitem{UPY18}
\bibinfo{author}{Irek \surnamestart Ulidowski\surnameend},
  \bibinfo{author}{Iain \surnamestart Phillips\surnameend} \&
  \bibinfo{author}{Shoji \surnamestart Yuen\surnameend} (\bibinfo{year}{2018}):
  \emph{\bibinfo{title}{Reversing Event Structures}}.
\newblock {\slshape \bibinfo{journal}{New Generation Computing}}
  \bibinfo{volume}{36}(\bibinfo{number}{3}), pp. \bibinfo{pages}{281--306},
  \doi{10.1007/s00354-018-0040-8}.

\bibitemdeclare{article}{GP09}
\bibitem{GP09}
\bibinfo{author}{Robert \surnamestart {van Glabbeek}\surnameend} \&
  \bibinfo{author}{Gordon \surnamestart Plotkin\surnameend}
  (\bibinfo{year}{2009}): \emph{\bibinfo{title}{Configuration Structures, Event
  Structures and Petri Nets}}.
\newblock {\slshape \bibinfo{journal}{Theoretical Computer Science}}
  \bibinfo{volume}{410}(\bibinfo{number}{41}), pp. \bibinfo{pages}{4111--4159},
  \doi{10.1016/j.tcs.2009.06.014}.

\bibitemdeclare{inproceedings}{W89}
\bibitem{W89}
\bibinfo{author}{Glynn \surnamestart Winskel\surnameend}
  (\bibinfo{year}{1989}): \emph{\bibinfo{title}{An Introduction to Event
  Structures}}.
\newblock In \bibinfo{editor}{J.~W. \surnamestart de~Bakker\surnameend},
  \bibinfo{editor}{W.~P. \surnamestart de~Roever\surnameend} \&
  \bibinfo{editor}{G.~\surnamestart Rozenberg\surnameend}, editors: {\slshape
  \bibinfo{booktitle}{Linear Time, Branching Time and Partial Order in Logics
  and Models for Concurrency}}, \bibinfo{publisher}{Springer Berlin
  Heidelberg}, \bibinfo{address}{Berlin, Heidelberg}, pp.
  \bibinfo{pages}{364--397}, \doi{10.1007/BFb0013026}.

\end{thebibliography}
\end{document}